\documentclass{aa}  

\usepackage{graphicx}
\usepackage[usenames,dvipsnames]{color}
\usepackage{xcolor}
\usepackage{enumitem}
\usepackage{float}
\usepackage{cuted}
\usepackage{soul}
\usepackage{scalerel}
\usepackage{tikz}
\usetikzlibrary{svg.path} 

\usepackage{hyperref}
\usepackage{txfonts}
\definecolor{orcidlogocol}{HTML}{A6CE39}
\tikzset{
  orcidlogo/.pic={
    \fill[orcidlogocol] svg{M256,128c0,70.7-57.3,128-128,128C57.3,256,0,198.7,0,128C0,57.3,57.3,0,128,0C198.7,0,256,57.3,256,128z};
    \fill[white] svg{M86.3,186.2H70.9V79.1h15.4v48.4V186.2z}
                 svg{M108.9,79.1h41.6c39.6,0,57,28.3,57,53.6c0,27.5-21.5,53.6-56.8,53.6h-41.8V79.1z M124.3,172.4h24.5c34.9,0,42.9-26.5,42.9-39.7c0-21.5-13.7-39.7-43.7-39.7h-23.7V172.4z}
                 svg{M88.7,56.8c0,5.5-4.5,10.1-10.1,10.1c-5.6,0-10.1-4.6-10.1-10.1c0-5.6,4.5-10.1,10.1-10.1C84.2,46.7,88.7,51.3,88.7,56.8z};
  }
}

\newcommand\orcidicon[1]{\href{https://orcid.org/#1}{\mbox{\scalerel*{
\begin{tikzpicture}[yscale=-1,transform shape]
\pic{orcidlogo};
\end{tikzpicture}
}{|}}}}

\begin{document} 

\def \foo XMM{XMM-\textit{Newton}}
\def \f 4U{4U\,1624--490}
   \title{An \foo XMM long look at the accretion disk plasma \\
   in the dipping neutron star LMXB \f 4U}
   \subtitle{}

\author{E. Caruso\inst{1,2\orcidicon{0009-0002-6843-7076}}, E. Costantini\inst{2\orcidicon{0000-0001-8470-749X}}, N. Degenaar\inst{1\orcidicon{0000-0002-0092-3548}}, M. D{\'i}az Trigo\inst{3\orcidicon{0000-0001-7796-4279}}}

   \institute{Anton Pannekoek Institute for Astronomy, University of       
            Amsterdam, Science Park 904, NL-1098 XH Amsterdam, the Netherlands \and
             SRON Space Research Institute Netherlands, Niels Bohrweg 4, NL-2333 CA Leiden, the Netherlands \and
             ESO, Karl-Schwarzschild-Strasse 2, D-85748 Garching bei München, Germany \\  
             }

   \date{Received June 25, 2025; accepted October 2, 2025}

 
  \abstract
   {Dipping neutron star low-mass X-ray binaries (NS LMXBs) are systems that exhibit periodic drops in their X-ray light curves. These are believed to be caused by material at the impact point of the gas stream onto the accretion disk, the bulge.
   Dipping systems are observed at high inclination and provide exceptional opportunities to address important open questions about accretion disks, such as the physical properties of the bulge, and the connection between disk atmospheres and disk winds.} 
   {We aimed to characterize the accretion disk plasmas present in the 21h-period NS LMXB \f 4U, and perform a detailed spectral analysis of the material present at the impact region.} 
   {We used four \foo XMM EPIC pn observations that were specifically targeting dips, and allow us to probe dipping activity over different timescales (i.e. consecutive orbits and $\sim$6 months). We use both time- and flux-resolved spectroscopic analysis to probe the structural properties of the bulge moving along the line of sight and its homogeneity, respectively.} 
   {During dipping, the primary spectrum is modulated by an ionized (log$\xi\sim$ 3.4) absorber with varying column density and covering factor, as well as a colder absorber. This suggests that the bulge is a multiphase and clumpy absorbing medium.
    From size scale arguments, we estimate the number of clumps in the bulge to be $>$7$\times10^{3}$.
    A highly ionized disk atmosphere becomes evident only when different phases of absorption are analyzed individually.
    This work demonstrates the feasibility of constructing a physical picture of the bulge, and highlights how future research could reveal how its properties depend on system parameters, and whether the bulge could influence the dynamics of the accretion disk.
    }
   {}


\titlerunning{An \foo XMM long look at the accretion disk plasma in the dipping neutron star LMXB \f 4U}
\authorrunning{E. Caruso et al}

   \maketitle
%

\section{Introduction}

Accretion of matter is a fundamental physical process that plays a role at various spatial scales in the universe, from the formation of stars and planets to the evolution of binary star systems and galaxies.
In low-mass X-ray binaries (LMXBs), the gas in the accretion disk is ionized by the strong X-ray emission from the region near the compact object, leading to the formation of photoionized plasmas. 
These plasmas can be observed as static (e.g. disk atmospheres and coronae) or outflowing (e.g. disk winds) with a certain velocity \citep{1983Begelman, Jimenez-Garate2002}. Such outflows carry away both angular momentum and mass, with implications on the accreting systems' evolution and their surrounding environment \citep[e.g][]{Ponti2012, Justham2012,2014Degenaar,2024Gallegos-Garcia}. 

LMXBs observed at high inclination are ideal laboratories for studying these accretion disk plasmas, which are detected through the presence of absorption lines from highly ionized elements in the X-ray spectra. Absorption from photoionized plasmas was first observed in the spectrum of the black hole LMXB GRO J1655-40 \citep{Ueda1998}. Since then, the powerful grating spectrographs on board of \foo XMM \citep{DenHerder_XMM_RSG2001} and \textit{Chandra} \citep{Canizares_Chandra2005}, have significantly increased the number of absorbers and spectral features detected. 
To date, evidence for the presence of ionized absorbers has been found in a total of 33 LMXBs, 19 neutron star and 14 black hole systems \citep{Neilsen_Degenaar2023}. The properties of these spectral features yield information on the location and physical condition of the absorbing plasmas in the accretion disk environment \citep[e.g][]{Xiang2009, Psaradaki2018, Trueba2020}. Therefore, studies focused on ionized absorbers in high inclination systems can help answer open questions related to accretion disk plasmas, such as understanding their geometry and location in the disk or investigating how and if static and outflowing plasmas are connected with each other.

Evidences of high inclination LMXBs are dipping and/or eclipsing phenomena \citep{Frank1987}. About two dozen dipping sources are known to date, with about half of them showing eclipses \citep{Avakyan2023}. Dipping phenomena result from the interaction between the turbulent and dense stream of material from the companion star and the impact region of the accretion disk. The material accumulating in the impact region periodically obscures the central X-ray source \citep{White1982, VanPeet2009, Boirin2005, DiazTrigo2006}. These interactions can significantly increase the size of the accretion disk rim, generating a thick bulge \citep{White1985}. Except for the debris left from the aftermath of the accretion stream impact, reported for LMXB EXO 0748-676 in \cite{Psaradaki2018}, to date, the specifics of the structure and plasma properties of this impact bulge region of the accretion disk have not been extensively investigated. 

The properties of the dips (e.g. duration, intensity, and substructures) vary significantly from source to source, and different models were proposed to explain the complex spectral changes observed between dipping and persistent (i.e. non-dipping) spectra. 
\cite{Parmar1986} modeled the spectrum in terms of a persistent, unabsorbed component with an associated absorbed one, characterized by a varying degree of opacity.
\cite{Church1997} explained dipping with the partial and progressive covering of a point-like source emission by an extended absorber. 
Finally, \cite{Boirin2005} developed a third model focused on ionized absorbers, were all spectral changes observed were modeled by variations in the properties of the ionized absorbers (e.g. column density and ionization). This model proved to be a viable interpretation for most of the known dipping sources \citep{DiazTrigo2006}.

\f 4U, also known as the ``Big Dipper'', is a bright - L$_{(0.6 - 10 keV)}$ $\sim$ 4.9 $\times$ $10^{37}$erg s$^{-1}$ \citep{Xiang2009} - dipping neutron star LMXB and an excellent target to study accretion disk and dipping plasmas. The orbital period of $\sim 21$ hr \citep{Jones1989} is the longest one amongst dipping sources showing regular dips, and it results in a larger separation between the compact object and the companion star, and larger accretion disk radius  \citep{Smale&Church2001} compared to other dipping sources. In addition, the 
long dipping activity \citep[$\sim 3$ hr;][]{Watson1985} makes the source ideal for the study of dips. 

The source was first discovered by EXOSAT \citep{Watson1985}. The hydrogen column density in the line of sight of \f 4U is substantial (6 -- 8  $\times10^{22}$ $\mathrm{cm}^{-2}$) and the source was found to have a prominent scattering halo component \citep{Angelini1997}. Dust scattering halos form when part of the source radiation is scattered by interstellar dust, and then directed back into the observer’s line of sight, with a certain delay. The response time the scattering halo of 4U 1624-490 has been estimated to be 1.6 ks \cite[$\sim$ 27 min;][]{Xiang&Lee&Nowak2007}.
The halo component has been shown to be essential for modeling of the continuum, especially during dipping \citep{DiazTrigo2006}.

The source is at a distance of $\sim$15 kpc, calculated using the halo response time \citep{Xiang&Lee&Nowak2007}, consistent with the value calculated by \cite{Christian&Swank1997}, which compared the hydrogen column densities derived from spectral fitting to models of exponential distribution of hydrogen in the Galaxy.

To date, the source has been studied using EXOSAT, \foo XMM, and \textit{Chandra}. Data from 1985 were first modeled with the complex continuum model \citep{Church1995}, then evidence of absorption of Fe\textsc{xxv} and Fe\textsc{xxvi} was found in \foo XMM data taken in 2001 \citep{Parmar2002}. These features were observed in both persistent and dipping spectra, but found to become shallower during the obscuration phase, suggesting the presence of an absorber less strongly ionized during dipping compared to the persistent phase. 
Further analysis of the same data set showed that all the spectral changes observed could be accounted for by a change in the properties of the ionized absorbers which, during dipping, increased in column density and decreased in ionization \citep{DiazTrigo2006}. 

\textit{Chandra} HETGS data taken in 2004 confirmed the presence of highly ionized iron and pointed at the presence of two different plasmas in the accretion disk from which these lines originated: a hot corona and a warm (i.e. less ionized) absorber located at the accretion disk rim \citep{Xiang2009}. However, only analysis of 'near dip' events \citep[Fig 1 in][]{Xiang2009} was included and dips were not analyzed. 

In this work, we use unpublished \foo XMM data and carry out the first detailed spectral analysis focused on the dipping behavior of the source, as well as a simultaneous analysis of dipping and persistent spectra. Specifically, we combine both a time-resolved and flux-resolved analysis in order to characterize the accretion disk plasmas present in  \f 4U and understand the structure of the turbulent impact region of the accretion stream, the bulge.

\section{Data analysis}

    The X-ray telescope on board of the \foo XMM space observatory \citep{Jansen_XMM2001} is equipped with two X-ray instruments, the European Photon Imaging Camera (EPIC) and the Reflection Grating Spectrometer \citep[RGS,][]{DenHerder_XMM_RSG2001}. The EPIC has a total of three cameras with two different types of CCD, two MOS \citep{Turner_XMM_MOS2001} and one pn \citep{Struder_XMM_PN2001}. The high hydrogen column density in the line of sight of 4U 1624 -- 490 causes a large amount of absorption at lower energies, making RGS data unavailable for the analysis.

    \begin{table}
      \caption[]{Observation log for the EPIC pn observations. }
         \label{obs_log}
         \centering
         \resizebox{7.5cm}{!}{\begin{tabular}{c c c c c} 
            \hline
            \hline
            \noalign{\smallskip}
            \textbf{Obs ID} & \textbf{Start date} &  \textbf{Exposure} & \textbf{Count rate}\\
            & (yyyy/mm/dd) & (ks) & (counts/s)\\            
            \noalign{\smallskip}
            \hline
            \noalign{\smallskip}
            0402330\textbf{301} & 2006-08-17 & 42.0 & 80.6 $\pm$ 5.3\\
            0402330\textbf{401} & 2007-02-10 & 22.4 & 60.2 $\pm$ 6.7\\
            0402330\textbf{501} & 2007-02-13 & 23.6 & 58.0 $\pm$ 6.3\\
            0402330\textbf{601} & 2007-02-15 & 25.3 & 63.1 $\pm$ 6.5\\
            \noalign{\smallskip}
            \hline
        \end{tabular}}
   \end{table}

   \begin{figure*}
   \centering
      \includegraphics[width=\textwidth]{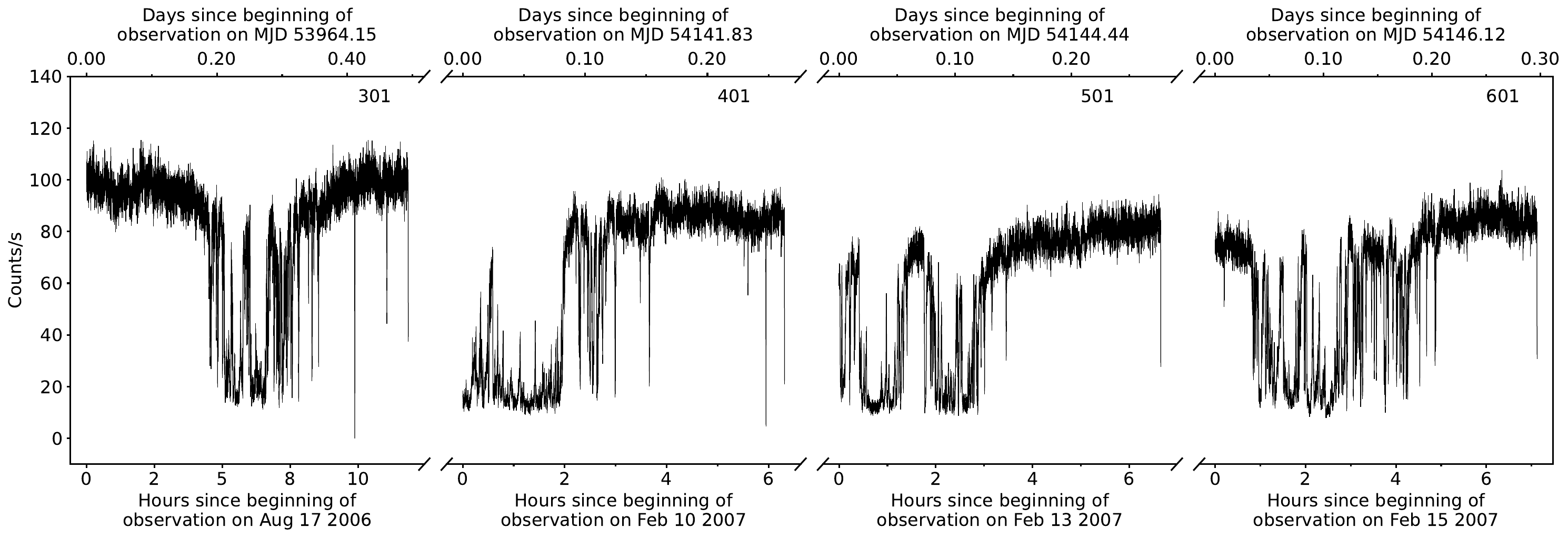}
      \caption{Composite 0.3 - 10 keV light curves of the four EPIC pn observations, with binning in time of 5s. The light curves show both persistent and dipping intervals.}
        \label{fig: lc_all}
   \end{figure*}

\begin{figure}
   \centering
      \includegraphics[width=0.40\textwidth]{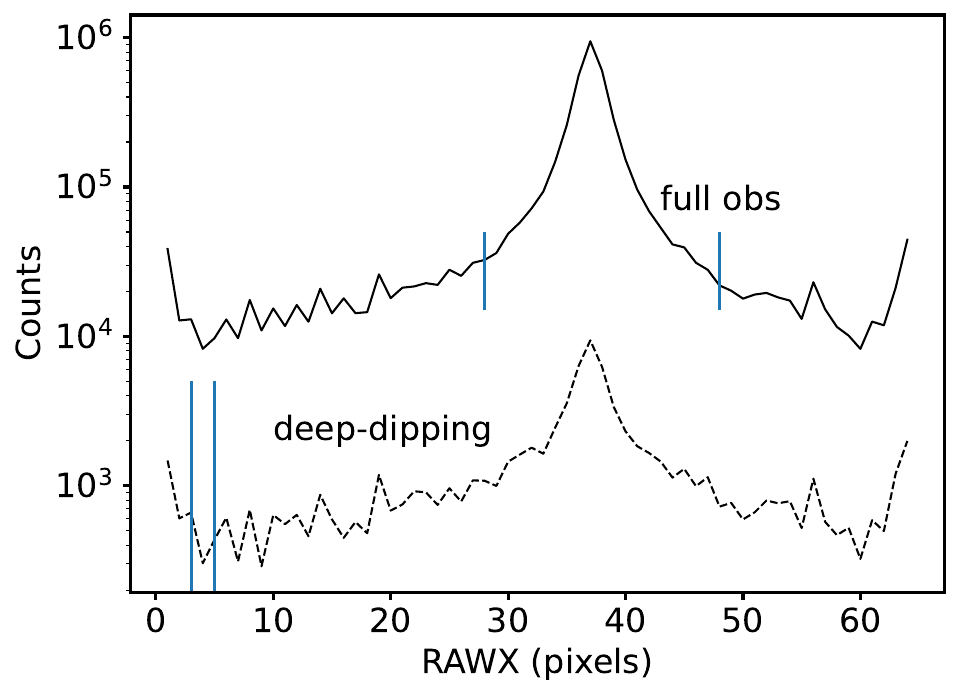}
      \caption{Histogram of the total number of photons counts vs RAWX for the original event file of the full observation (solid line) and for the one from the \textit{deep-dipping} time interval (dotted line). The defined source region RAWX = [28 -- 48] and background region RAWX = [3 -- 5] are indicated with vertical lines. The plot shows the background contamination by the source emission displayed by the difference in counts around the wings of the distributions. Data shown for obs 301.}
        \label{fig: phtcnt_vs_pixels}
   \end{figure}

    Table \ref{obs_log} shows the log of the observations used in this work. All observations targeted the dipping phase of the source, and the corresponding light curves are shown in Fig \ref{fig: lc_all}. This dataset allows the study of the dipping phenomena over different timescales, as obs 401 was taken $\sim$ 6 months after obs 301, while obs 401 to 601 belong to consecutive orbits. The observations were taken with the pn camera in timing mode, which is used to increase time resolution (up to 30 $\mu s$) and avoid pile-up in the data. With this observation mode, the data on the CCD are collapsed into a one-dimensional row along the y-axis. Spatial information is maintained only along the x-axis (RAWX), which indicates the pixels of the pn CCD chip. 
    The data were retrieved from the \foo XMM public archive and all data products were extracted using the Science Analysis Software (SAS, v.21.0.0) and HEASOFT (v.6.32). We verified that the observations were not affected by background flares. To select a source and background region, we looked at the distribution of photon counts over the CCD. For all observations, we defined the source region to be RAWX = [28 -- 48] (Fig. \ref{fig: phtcnt_vs_pixels}).  
    
    For timing mode observations of bright sources, the edges of the CCD can still be significantly contaminated by the source emission and/or scattering halos\footnote{\scriptsize\textit{\foo XMM Calibration Technical Note}, XMM-SOC-CAL-TN-0083}.  For the background region, we initially selected RAWX = [3 -- 5]  however, for all observations the background was highly contaminated with source emission. 
    To minimize this effect, we selected the time interval that included the deepest of the dipping activity - which represents the period with the least amount of radiation coming from the source, and therefore the least contaminated background available, as shown in Fig \ref{fig: phtcnt_vs_pixels}. Therefore, we selected the background region to be again RAWX = [3 -- 5], but extracted from the \textit{deep-dipping} interval event file (Fig \ref{fig: phtcnt_vs_pixels}).
    We scaled this new background file to the correct exposure for each of the time intervals selected for the analysis ($\S$~\ref{subsec: time intervals}). This was done through the \texttt{fakeit} command in XSPEC v. 12.13.1\citep{1996XSPEC}. A simple absorbed powerlaw was provided as input model to \texttt{fakeit}.
    This background extraction procedure was carried out for each observation individually (i.e. the background file was extracted from the \textit{deep-dipping} of each observation). Finally, we verified that the data were not affected by pile up, using \texttt{epatplot}.

    \section{Spectral analysis and modeling}

   \begin{figure*}
   
   \centering
   \includegraphics[width=13.5cm,angle=0]{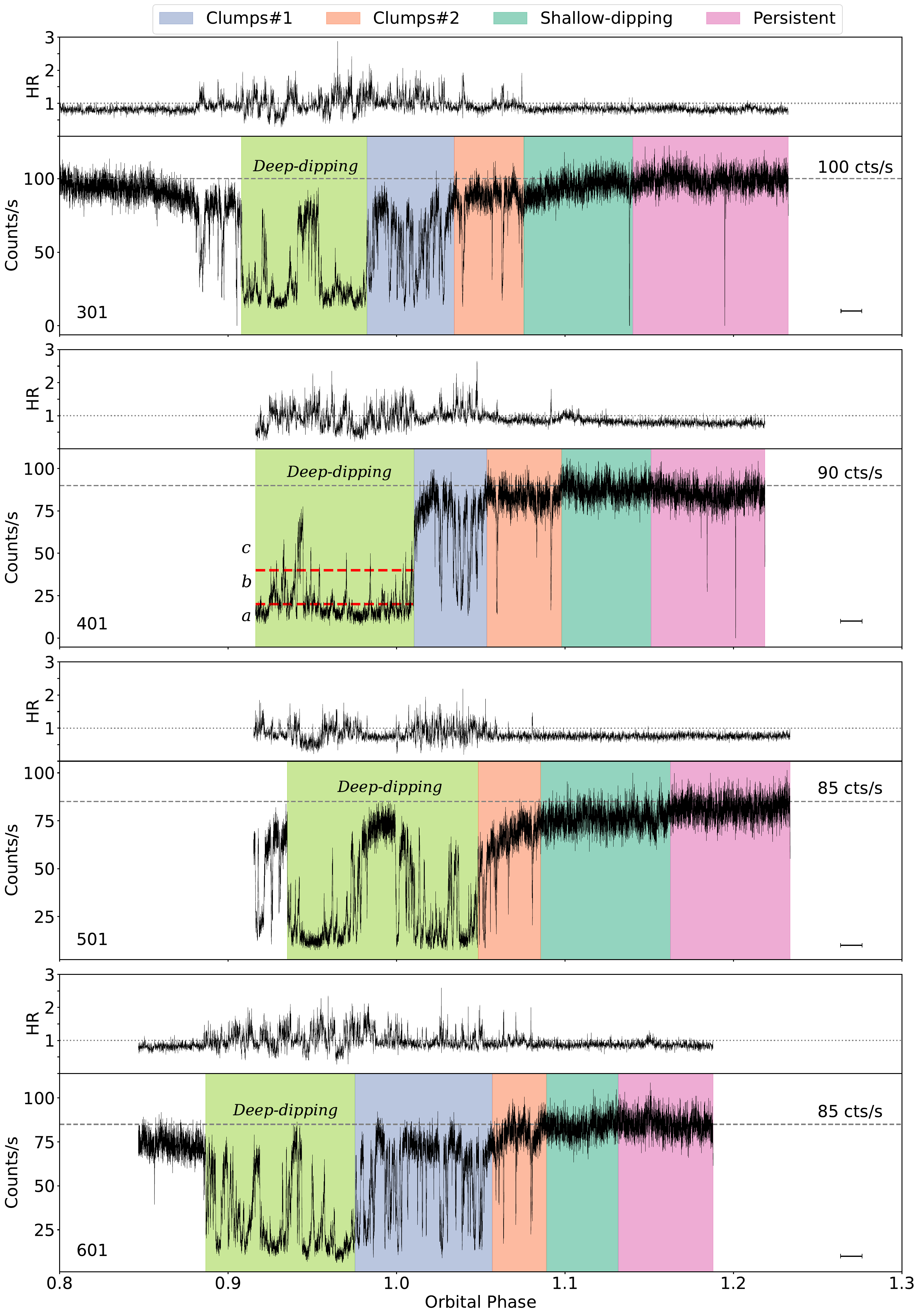}
   \caption{Phase folded light curves of all EPIC pn observations based on the ephemeris from \cite{Smale2001} - with phase 0 defined as the centre of the dip. The uncertainty on phase due to error propagation of the ephemeris is displayed as an error bar in the bottom right corner of each plot. All light curves are 0.3 -- 10 keV and have a 5s time binning. In the top panel of each plot the hardness ratios (5 -- 10 keV/1.7 -- 5 keV) for each observation are shown. The average persistent count rate is indicated as a dashed line in each light curve. The time intervals selected for the time-resolved analysis are shown for all observations: \textit{clumps$\#$1}, \textit{clumps$\#$2}, \textit{shallow-dipping}, and \textit{persistent} (light blue, orange, dark green, and pink colors, respectively - also see Section $\S$\ref{subsec: time intervals}). The duration of the \textit{deep-dipping} interval is shown in all observations (light green). The intensity ranges for the flux-resolved analysis of the \textit{deep-dipping} in obs 401 are also indicated (\textit{a}, \textit{b}, \textit{c}) - see Section $\S$\ref{sub: deep dip}.}
   \label{fig: lc_fold_int}
   \end{figure*}

    \subsection{Light curves and time intervals}
    \label{subsec: time intervals}
    
     Dips are present in all observations and a great variety of light curve profiles are displayed (Fig \ref{fig: lc_all}). Obs 301 and 601 cover the start of the dipping activity, while the source was already dipping when obs 401 and 501 began. The bulk of the dipping and the post-dipping activity are sampled by all four observations. Using the ephemeris calculated by \cite{Smale2001}, we phase-folded the data to investigate the recurrence and possible nature of the modulations observed. The 0.3 -- 10 keV phase-folded light curves are shown in Fig \ref{fig: lc_fold_int} with a time binning of 5s.

    The quality of the data sets and the chosen time binning unveil types of modulations that seem to be recurrent. In all observations, it is possible to identify the bulk of the dipping coverage followed by an irregular and gradual flux recovery as the bulge moves away from the line of sight. We classified the variety of modulations observed in the light curves as described below:

    \textit{Deep-dipping}: the heaviest part of the obscuration due to dipping activity (Fig. \ref{fig: lc_fold_int}). Throughout the observations, it displays a similar duration of $\sim$2 hours (around 10\% of the orbital period). The obscuration shows non-uniform coverage that allows the radiation to temporarily pass through. This results in an apparent W shape in the light curve. Dipping activity occurs during the entire interval duration;
    
    \textit{Clumps$\#$1}: very short, nearly fully absorbed intervals that remind of 'sporadic' clumps passing through the line of sight (hence the name). During these intervals, the observed count rate drops to values very close to those of the \textit{deep-dipping}, but there are changes in the modulation observed, namely its duration and shape, distinct from the \textit{deep-dipping}. Dipping activity is always at least $>1/2$ of the interval duration;
        
    \textit{Clumps$\#$2}: with these intervals we sample the modulation where the flux recovery from the dipping is interspersed with a lower amount of sudden deep coverage shown by fast drops in count rate. In this cathegory, the dipping activity is around $\le 1/3$ of the interval duration;
    
    \textit{Shallow-dipping}: intervals where we sample a gradual and constant flux recovery; 
    
    \textit{Persistent}: intervals where no modulation is apparent;

    These selections are not merely based on a flux variation, as in previous studies \citep[e.g.][]{DiazTrigo2006, Boirin2005}, but they are meant to point at different time-varying structures that contribute to the obscuration observed.
    These types of modulations are identified in all observations, with only a few exceptions. For example, in obs 401 and 601, the gradual count rate recovery during the \textit{shallow-dipping} is not as evident as in obs 301 and 501. The \textit{clumps$\#$1} interval does not appear in obs 501. In obs 601, its seems like the \textit{clumps$\#$1} interval has doubled in duration (Fig \ref{fig: lc_fold_int}). 
    
    We used these intervals to perform a detailed time-resolved spectral analysis. We aimed to systematically investigate the spectral properties of the plasmas present throughout the dipping phase and their evolution in time. The selected categories are also shown in Fig \ref{fig: lc_fold_int}.
    The \textit{deep-dipping} was excluded from the time-resolved analysis, as it shows a variety of profiles and a level of contamination (i.e. different amounts of source radiation passing through the blocking medium) that preclude a systematic study throughout all observations. This is also shown by the HR diagrams (Fig \ref{fig: lc_fold_int}; note that for the HR we ignore data points below 1.7 keV - as we do during spectral fitting - due to the high absorption). Spectral hardening has been observed during dipping activity \citep{DiazTrigo2006, Boirin2005} and it is clearly visible in the \textit{clumps$\#$1} and \textit{clumps$\#$2} intervals. However, during the \textit{deep-dipping} intervals in all observations, the variations in the HR are quite complex and structured, showing a combination of hardening and softening. Therefore, for the analysis of the \textit{deep-dipping} intervals we instead performed a flux-resolved analysis, described $\S$ \ref{sub: deep dip}.


    \subsection{Baseline model}
    \label{sub: model}
\begin{figure}
    \centering
      \includegraphics[width=0.45\textwidth]{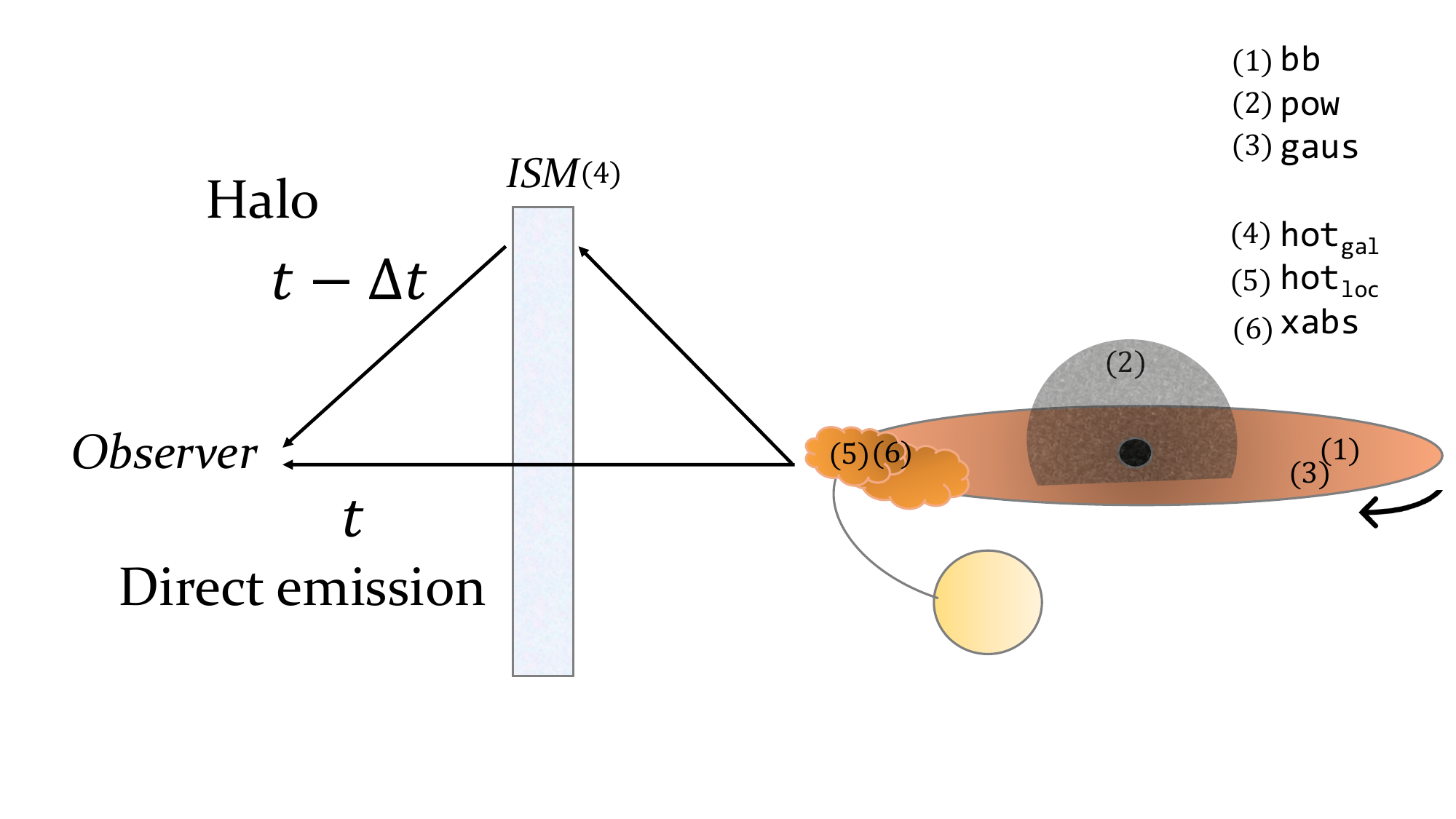}
      \caption{Schematic of our employed spectral model. The sketch illustrates the direct source emission coming from a point in time \textit{t}, and the halo - delayed source emission from the moment \textit{t - $\Delta$t}. A legend for the model components (SPEX syntax) used is also shown.}
        \label{fig: model}
   \end{figure}

    We performed spectral analysis using SPEX version 3.08.01 \citep{1996spex}. All data were optimally rebinned within SPEX \texttt{obin}. Uncertainties on spectral fitting parameters are calculated with the \texttt{error} command and reported as 1$\sigma$ errors, corresponding to a 68$\%$ confidence level. 
    We constructed a baseline model consisting of the accretion flow, the absorption components, and the scattering halo. 
    The continuum accretion flow components are the disk (or neutron star emission/boundary layer), the corona, and the fluorescent iron line coming from the disk reflection (see Fig \ref{fig: model}, components 1-3). These are described with a black body (\texttt{bb}), a powerlaw (\texttt{pow}), and a Gaussian (\texttt{gaus}) components, respectively.
    
    The absorbing components that attenuate the accretion flow emission are the constant Galactic absorption and a time-varying absorption local to the system, caused by the bulge structure moving around with the disk (see Fig \ref{fig: model}, components 4-6).
    Galactic absorption is modeled with a collisionally ionized absorber (\texttt{hot$_{gal}$}). The time-varying local absorption has both a cold and a ionized components \citep{DiazTrigo2006, Xiang&Lee&Nowak2007, Xiang2009}, and therefore we use an additional cold component (\texttt{hot$_{loc}$}) together with a photoionized absorber component (\texttt{xabs}).

    To account for the expected scattering halo, we followed the model previously defined by \cite{DiazTrigo2006}. We first reduce the source emission (i.e. absorbed accretion flow emission) by a factor of e$^{-\tau(\mathrm{E})}$ (\texttt{etau} model in SPEX), which corresponds to the scattering of photons out of the line of sight. The optical depth $\tau(\mathrm{E})$ is defined as $\tau_{\mathrm{1}}$E${^\mathrm{-2}}$, where $\tau_{\mathrm{1}}$ is the optical depth at 1 keV and E${^\mathrm{-2}}$ the energy dependence for the dust-scattering cross-section. Throughout the text, we refer to this first part of the model as "direct emission". We then added a second part to our model where the source emission is multiplied by a factor (1 - e$^{-\tau(\mathrm{E})}$), which accounts for that part of the halo photons that are scattered back into the line of sight (see Fig \ref{fig: model}). Similarly to the first part of the model, a \texttt{hot$_{gal}$} and \texttt{hot$_{loc}$} components are used for cold absorption from the ISM and local to the source, respectively. Limited by the data quality, no ionized absorption local to the source is included in the second part of the model. A sketch of our model is shown in Fig~\ref{fig: model}, and our complete model in SPEX syntax is: 
    
    \smallskip
    \begin{centering}
        \texttt{etau}$\ast$[\texttt{hot$_{gal}$}$\ast$\texttt{hot$_{loc}$}$\ast$\texttt{xabs}(\texttt{bb}+\texttt{pow}+\texttt{gaus})]+ \\
        (1- \texttt{etau})$\ast$[\texttt{hot$_{gal}$}$\ast$\texttt{hot$_{loc}$}$\ast$(\texttt{bb}+\texttt{pow}+\texttt{gaus})] \\
    \end{centering}

    \smallskip

   Our aim is to investigate the time-variable absorbing components related to the bulge structure. 
   Therefore, following the approach proposed by \cite{Boirin2005} and \cite{DiazTrigo2006}, we kept the baseline model fixed and modeled all spectral changes during dipping intervals by allowing the properties (i.e. column density, ionization) of the cold and ionized absorbing components present in the accretion disk to vary.
   Specifically, all continuum parameters were coupled and the absorbers' properties were allowed to change throughout the time intervals. Note that with this approach we assume that the accretion flow emission stays constant, exceptions are reported in Appendix \ref{subsec: model complex}. For each observation, we used the baseline model and fitted all the spectra from the different time intervals simultaneously. 
    
    Further fitting details of the baseline model:
    \setlist{nolistsep}
    \begin{itemize}[noitemsep]    
    \item The medium temperature was fixed to 10$^{\mathrm{-6}}$ keV to mimic a cold absorber for both Galactic and local cold absorption;
    \item The Galactic column density $N_{\mathrm{H}}$ was fixed to 2.2 $\times10^{22}$ $\mathrm{cm}^{-2}$ \citep[][]{HI4PI2016};
    \item For photoionized absorption, the column density $N_{\mathrm{H}}$ and ionization level $\xi$ were allowed to change. \citet{Xiang2009} reported an upper limit on the outflow velocity of the plasma (263 $\mathrm{km/s}$). Given EPIC pn's resolution, the outflow velocity of the absorbers was fixed to zero. 
    \end{itemize}

 \begin{figure}
   \centering
   \includegraphics[width=9cm,angle=0]{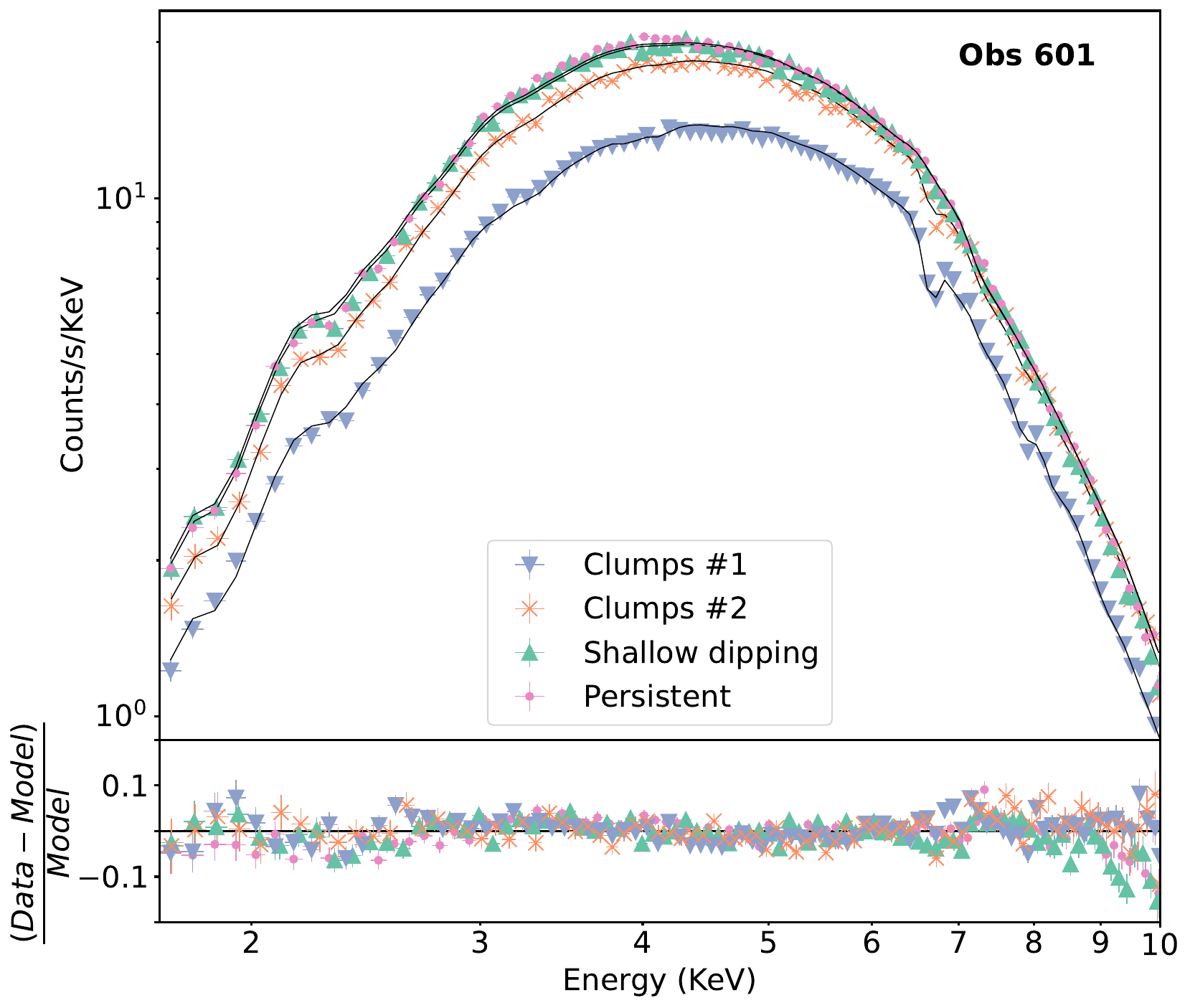}
   \caption{Spectra from the selected time intervals of obs 601 with the same color-coding as Fig \ref{fig: lc_fold_int}. Black solid lines represent the model for each spectrum, and residuals are shown in the bottom panel.}
   \label{fig: spec_601}
   \end{figure}

    \begin{table*}
      \caption[]{Best fit parameters and errors for the time-resolved spectral results obs 601. }
         \label{tab: res_601}
         \centering
         \resizebox{11.5cm}{!}{\begin{tabular}{c c c c c c c} 
            \hline
            \hline
            \noalign{\smallskip}
            \textbf{Obs 601} & & & Clumps$\#$1 & Clumps$\#$2 & Shallow & Persistent\\
            \noalign{\smallskip}
            \hline
            \noalign{\smallskip}
            \small{Comp} & \small{Param} & \small{Unit} & & & & \\
            & & & & & & \\
            \texttt{hot$_{gal}$} & $N_{\mathrm{H}}$ & \small{($10^{\mathrm{22}}$ $\mathrm{cm}^{-2}$)} & & & & 2.2 (f)\\[1ex]
            \texttt{hot$_{loc}$} & $N_{\mathrm{H}}$ & \small{($10^{\mathrm{22}}$ $\mathrm{cm}^{-2}$)} & 11.1$\pm$0.1 & 9.64$\pm0.10$ & 8.70$\pm$0.04 & 8.50$\pm0.05$ \\[1ex]
            \texttt{bb}&  norm  & \small({$10^\mathrm{11}$} $\mathrm{cm}^{2}$) & & & & 57$_{-7}^{+9}$  \\[1ex]
            & \textit{T} & \small{(keV)} & & & & 1.46$\pm0.02$ \\[1ex]
            \texttt{pow} & norm & \small{($10^{44}$ ph/s/keV)} & & & & 73.5$_{-3.7}^{+0.8}$ \\[1ex]
            & $\Gamma$ & & & & & 1.91$_{-0.04}^{+0.01}$ \\[1ex]
            \texttt{gaus} & norm & \small{($10^{44}$ ph/s)} & & & & (5.5$\pm$1.4)$\times10^\mathrm{2}$\\[1ex]
            & \textit{E} & \small{(keV)} & & & & 6.5$\pm0.04$ \\[1ex]
            & FWHM & \small{(keV)} & & & & 0.23 (f) \\[1ex]
            \texttt{xabs} & $N_{\mathrm{H}}$ & \small{($10^{\mathrm{22}}$ $\mathrm{cm}^{-2}$)} & 30.4$\pm$0.6 & 3.6$_{-0.3}^{+0.5}$ &  & \\[1ex]
            & log $\xi$ & & 3.59$\pm$0.03 & 3.56$\pm$0.06 & / & / \\[1ex]
            & fcov & (\%) & $>$58 & $>$84 & & \\[1ex]
            \small{\textbf{Halo}} & & & & & &\\ [0.4ex]
            \texttt{hot$_{gal}$} & $N_{\mathrm{H}}$ & \small{($10^{\mathrm{22}}$ $\mathrm{cm}^{-2}$)} & & & & 2.2 (f)\\[1ex]
            \texttt{hot$_{loc}$}& $N_{\mathrm{H}}$ & \small{($10^{\mathrm{22}}$ $\mathrm{cm}^{-2}$)} & 8.0$\pm0.1$ & 8.0 (f) & 8.0 (f) & 8.0 (f)\\[1ex]
            \texttt{etau} & tau0 & & & & & 2.50$_{-0.03}$ \\
            \noalign{\smallskip}
            \hline
            \noalign{\smallskip}
            $C$ (d.o.f) & 697 (344)& & & & \\
            \noalign{\smallskip}
            \hline
        \end{tabular}}
    \begin{minipage}{\textwidth}

    \vspace{0.1cm}


    \small Note: (f) parameter frozen to shown value; In the text, we refer to results on cold absorption as \texttt{hot$_{loc}$} + \texttt{hot$_{gal}$} as discussed in $\S$ \ref{sub: model}.
    All parameters are coupled when no value is shown, while "/" indicates that the component is not included.

    \end{minipage}
   \end{table*}

    \subsection{Time-resolved analysis: simultaneous spectral fitting from \textit{Clumps$\#$1} through \textit{Persistent}}
    \label{sub:time resolved}
    In this section, we report on the spectral results of our time-resolved analysis, which was performed for all observations and included all time intervals defined in Section $\S$ \ref{subsec: time intervals}, except the \textit{deep-dipping}.
     In the following, we use obs 601 to illustrate the main results of our time-resolved analysis and also discuss the general trends in the context of all the observations. Further details of the spectral results of obs 301, 401, and 501 are reported in Appendix \ref{app: A}. 
    
\begin{figure}
   \centering
   \includegraphics[width=0.37\textwidth,angle=0]{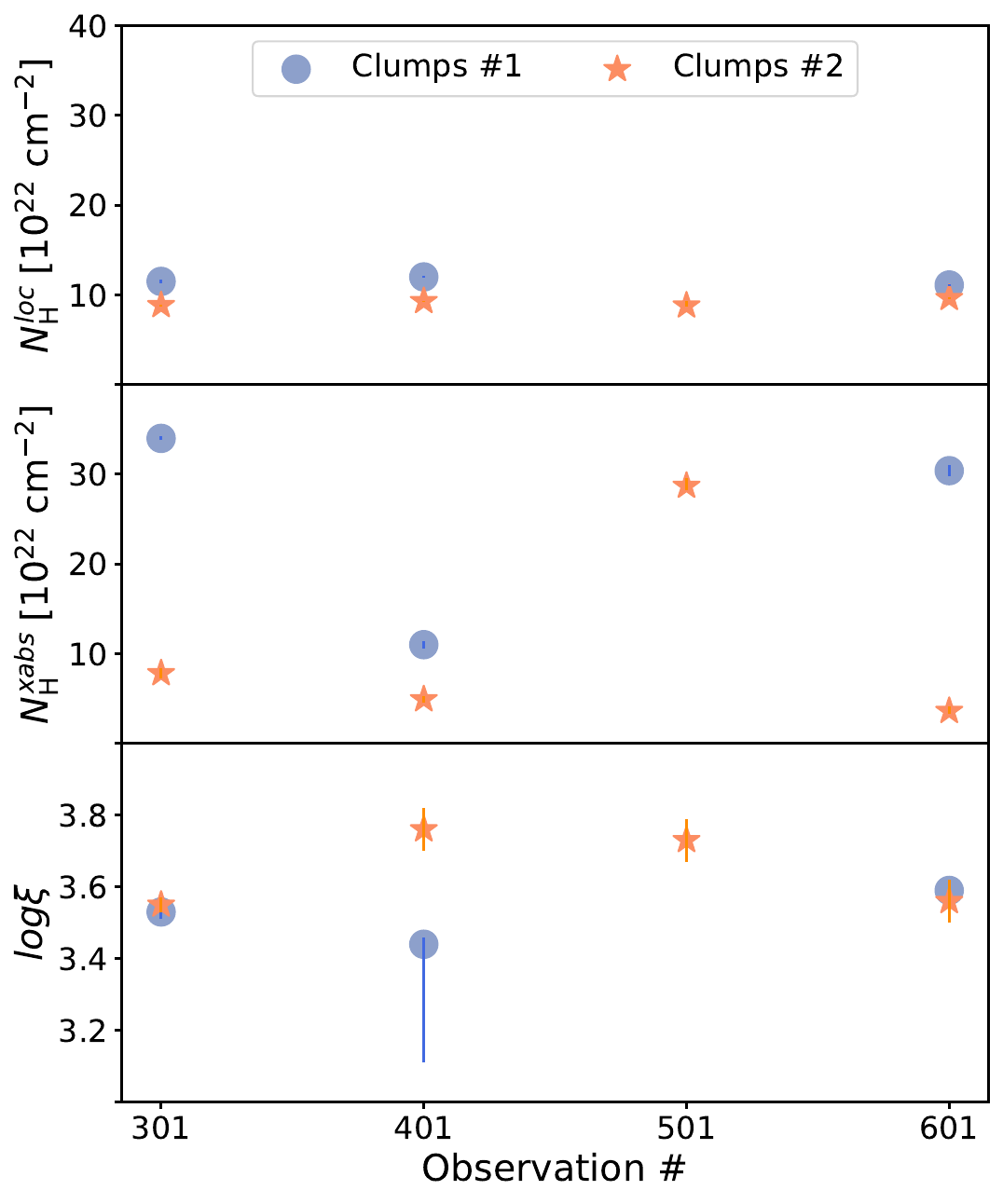}
   \caption{Trends from the spectral 
   results of the \textit{clumps\#1} and \textit{clumps\#2} intervals, for all observations (301, 401, 501, 601). There was no \textit{clumps\#1} identified for obs 501 (see $\S$ \ref{subsec: time intervals}).
   From top to bottom panels: column density value of local cold absorption ($N_{\mathrm{H}}^{loc}$), column density value of ionized absorption ($N_{\mathrm{H}}^{xabs}$), and ionization of the gas (log$\xi$). The uncertainty on values is shown with errorbars. If absent, the errors are too small to be visualized. }
   \label{fig: spec_trends}
   \end{figure}

    The spectral results of obs 601, including model and residuals are shown in Fig \ref{fig: spec_601}. Our model describes the spectra well, and the best-fit parameters are shown in Table \ref{tab: res_601}.  In the \textit{persistent} interval, the continuum accretion flow emission is described by a power law with index $\Gamma\sim$1.9, a black body component with a temperature T of $\sim$1.5 keV, and a Gaussian emission feature centered at 6.5 keV. We fixed the full width half maximum (FWHM) of the Gaussian feature to 0.23 keV, following \cite{Xiang2009}. 
    The spectra from the other intervals were coupled to these continuum parameters and all spectral differences are modeled solely by changes in cold and ionized absorption. We list and discuss the results of the selected time intervals:
    
    \setlist{nolistsep}
    \begin{itemize}[noitemsep]
    \item \textit{clumps$\#$1} and \textit{clumps$\#$2} (i.e. progressively moving towards the outer edge of the bulge): of the local time-varying absorption, the colder material (\texttt{hot$_{loc}$}) has the highest $N_{\mathrm{H}}$ value of $\sim$13$\times10^\mathrm{22}$ $\mathrm{cm}^{-2}$ in the \textit{clumps$\#$1} spectrum (i.e. closest to the \textit{deep-dipping}, Table \ref{tab: res_601}). 
    The column density then gradually starts decreasing from the \textit{clumps$\#$2} interval (i.e. as the bulge moves away from the line of sight). Ionized absorption (\texttt{xabs}) is detected in both intervals (Fig \ref{fig: spec_601}, light blue and orange spectra respectively). The absorbers have similar ionization value log$\xi$ $\sim$3.5, while the column density is a factor $\sim$10 larger in \textit{clumps$\#$1} compared to \textit{clumps$\#$2}. This hints to a significant decrease in the amount of absorbing material going from the former to the latter interval.
    
    In Fig \ref{fig: spec_trends} we plotted the properties (i.e column density and ionization level) of the cold and ionized gas present in \textit{clumps\#1} and \textit{clumps\#2} for all four observations. 
    We notice that local cold absorption in \textit{clumps\#1} and \textit{clumps\#2} (Fig \ref{fig: spec_trends}, top panel) have column density values ($N_{\mathrm{H}}^{loc}$) consistent through all observations.
    
    In terms of ionized absorption (Fig \ref{fig: spec_trends}, middle panel), the column density value ($N_{\mathrm{H}}^{xabs}$) for \textit{clumps\#1} shows some variations up to a factor 3. For \textit{clumps\#2}, the results are consistent through obs 301, 401, and 601. 
    The column density of the absorber is largest in obs 501, which could be explained by the additional slow and gradual attenuation of source emission on top of the sporadic clumps, displayed in the light curve (Fig \ref{fig: lc_fold_int}).
    
    The ionization of the gas (Fig \ref{fig: spec_trends}, bottom panel) appears to always have a value log$\xi\sim$3.5, with some small variations between observations\footnote{\scriptsize Note that obs 301, 401, and 601 have continuum parameters comparable within errors (see Table \ref{res_301}, \ref{res_401}, and \ref{tab: res_601}) which makes it for a meaningful comparison of the absorbers properties (e.g. log$\xi$) throughout \textit{clumps\#1} and \textit{clumps\#2}}.
    
    Overall the spectral results of \textit{clumps\#1} and \textit{clumps\#2} appear quite constant over time, which provides a picture on how the properties of the absorbing gas in the bulge have evolved over the time scales sampled (i.e. 301 vs 601, $\sim$6 months, and 401 vs 601, consecutive orbits i.e. 21h).
    Our results show that the dipping plasmas' properties do not seem to have changed over the time scale of months (obs 301 vs 601), although small changes are evident over consecutive orbits (401 vs 501 vs 601). 

    \item \textit{shallow-dipping} and \textit{persistent}: the local time-varying cold absorption (\texttt{hot$_{loc}$}) follows a decreasing trend from \textit{clumps$\#$1} to the \textit{persistent} spectra.
    This trend is consistent throughout all observations.
    In the \textit{persistent} interval, cold absorption local and Galactic always adds up to a column density value $N_{\mathrm{H}}$ $\sim$10.7$\times10^\mathrm{22}$ $\mathrm{cm}^{-2}$, which implies that a significant fraction is local to the source, as expected \citep[][]{DiazTrigo2006, Xiang&Lee&Nowak2007, Xiang2009}.

    In the \textit{shallow-dipping} and \textit{persistent} intervals, the ionized absorber was not significantly detected (Fig \ref{fig: spec_601}, green and pink spectra respectively) while the spectra were fitted simultaneously with the \textit{clumps$\#$1} and \textit{clumps$\#$2} intervals.
    This was the case for all observations (Table \ref{res_301}, \ref{res_401}, \ref{res_501} and Table \ref{tab: res_601}), and it is supported by the HR diagram showing no changes between the two intervals (Fig. \ref{fig: lc_fold_int} top panels).
    Nonetheless, we performed a separate additional analysis on all the \textit{shallow-dipping} and \textit{persistent} intervals individually, to further investigate the presence of ionized absorbers (see Table \ref{tab: pers_shall_table}, Appendix \ref{app: B}).

    \end{itemize}

   \begin{figure}
   \centering
   \includegraphics[width=9cm,angle=0]{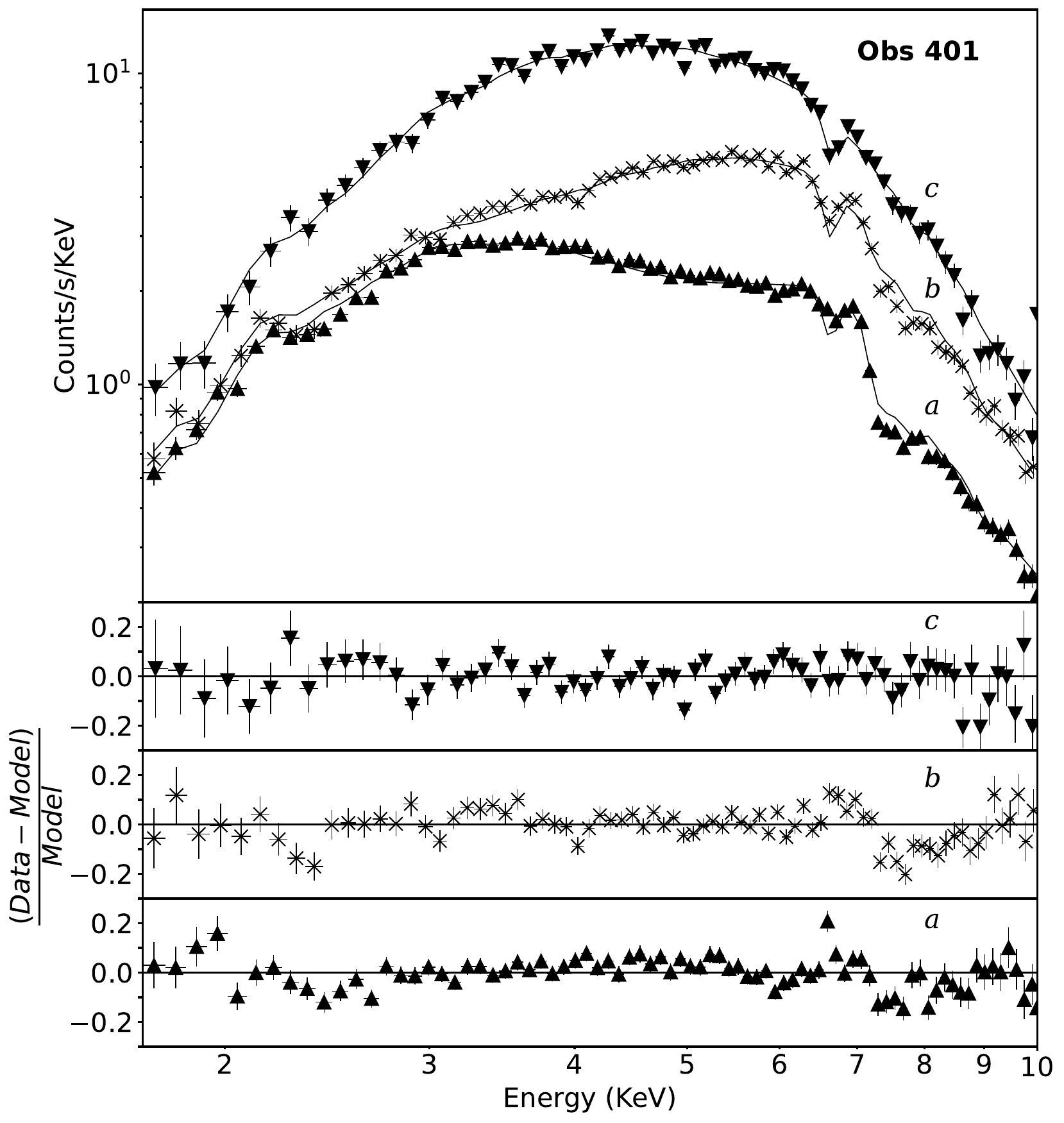}
   \caption{Spectra from the selected intensity intervals for the \textit{deep-dipping} of obs 401. The spectra are labeled following the intervals shown in the second panel of Fig \ref{fig: lc_fold_int}. Black solid lines represent the model for each spectrum, and residuals for each spectrum are shown in the bottom panels.}
   \label{fig: spec_401_fc}
   \end{figure}

    \begin{table*}
      \caption[]{Best fit parameters and errors for the flux-resolved spectral results obs 401.}
         \label{tab: res_401_fc}
         \centering
         \resizebox{12cm}{!}{\begin{tabular}{c c c c c c} 
            \hline
            \hline
            \noalign{\smallskip}
            \textbf{Obs 401} & & & \textit{a} : (cts/s $\le$ 20) & \textit{b} : (20 $<$ cts/s $\le$ 40) & \textit{c} : (cts/s $>$ 40)\\
            \noalign{\smallskip}
            \hline
            \noalign{\smallskip}
            \small{Comp} & \small{Param} & \small{Unit} & & & \\
            & & & & & \\
            \texttt{hot$_{gal}$} & $N_{\mathrm{H}}$ & \small{($10^{\mathrm{22}}$ $\mathrm{cm}^{-2}$)} & & & 2.2 (f)\\[1ex]
            \texttt{hot$_{loc}$} & $N_{\mathrm{H}}$ & \small{($10^{\mathrm{22}}$ $\mathrm{cm}^{-2}$)} & 82.8$_{-1.9}^{+0.5}$ & 34.0$\pm$0.7 & 9.7$\pm$0.5 \\[1ex]
            \texttt{bb}&  norm  & \small($10^\mathrm{11}$ $\mathrm{cm}^{2}$) & & & 61 (f) \\[1ex]
            & \textit{T} & \small{(keV)} & & & 1.41 (f)\\[1ex]
            \texttt{pow} & norm & \small{($10^{44}$ ph/s/keV)} & & & 72 (f) \\[1ex]
            & $\Gamma$ & & & & 1.88 (f)\\[1ex]
            \texttt{gaus} & norm & \small{($10^{44}$ ph/s)} & & & 7.3$\times10^{-2}$ (f)\\[1ex]
            & \textit{E} & \small{(keV)} & & & 6.46 (f) \\[1ex]
            & FWHM & \small{(keV)} & & & 0.23 (f)\\[1ex]
            \texttt{xabs} & $N_{\mathrm{H}}$ & \small{($10^{\mathrm{22}}$ $\mathrm{cm}^{-2}$)} & 81.6$\pm$3.4 & 60.0$_{-1.8}^{+0.7}$ & 48.0$\pm$(0.1$\times10^{-2})$ \\[1ex]
            & log $\xi$ & & & & 3.45 (UL)$^*$ \\[1ex]
            & fcov & (\%) & $>$57 & >57 & 87.0$\pm$(0.1$\times10^{-1})$ \\[1ex]
            \small{\textbf{Halo}} & & & & &\\ [1ex]
            \texttt{hot$_{gal}$} & $N_{\mathrm{H}}$ & \small{($10^{\mathrm{22}}$ $\mathrm{cm}^{-2}$)} & & & 2.2 (f)\\[1ex]
            \texttt{hot$_{loc}$}& $N_{\mathrm{H}}$ & \small{($10^{\mathrm{22}}$ $\mathrm{cm}^{-2}$)} & 9.30$\pm$0.05 & 9.0$\pm$0.1 & 9.0$\pm0.5$\\[1ex]
            \texttt{etau} & tau0 & & & & 2.5 (f) \\ [1ex]
            \noalign{\smallskip}
            \hline
            \noalign{\smallskip}
            $C$ (d.o.f) & 416 (222) & & &  & \\
            \noalign{\smallskip}
            \hline
        \end{tabular}}
    \begin{minipage}{\textwidth}

    \vspace{0.1cm}

    \vspace{0.1cm}

    \small Note: (f) parameter frozen to shown value. All parameters are coupled when no value is shown. In the text, we refer to results on cold absorption as \texttt{hot$_{loc}$} + \texttt{hot$_{gal}$} as discussed in $\S$ \ref{sub: model}. 

    \end{minipage}
   \end{table*}
   
    \subsection{Flux-resolved analysis: \textit{deep-dipping} 
    spectral results}
    \label{sub: deep dip}

    Here we summarize the analysis and results of the \textit{deep-dipping} i.e. when the accretion emission is most heavily obscured and we are looking at the most impenetrable part of the bulge.
    As mentioned in $\S$ \ref{subsec: time intervals}, the \textit{deep-dipping} intervals of the four observations showed different amounts of source radiation passing through the blocking medium, which does not make a time-resolved analysis straightforward. Therefore, we opted for a different approach and performed a flux-resolved spectral analysis of one of the \textit{deep-dipping} intervals. 
    We selected the one that showed the least amount of source radiation passing through, the \textit{deep-dipping} interval of obs 401, and extracted spectra based on dipping depth in the light curve. Specifically, a drop in count rate of $\sim$80\% (\textit{a}), between $\sim$80\% and $\sim$50\% (\textit{b}), and <$\sim$50\% (\textit{c}) (Fig \ref{fig: lc_fold_int}).
    With the flux-cut analysis, we study parts of the bulge that have maximal coverage such that almost all source emission is blocked (\textit{a}), to parts where the radiation can pass through sparsely (\textit{b}) or almost completely (\textit{c}). 
    For these flux intervals, we describe below their spectral results.

\begin{figure}
   \centering
   \includegraphics[width=7cm,angle=0]{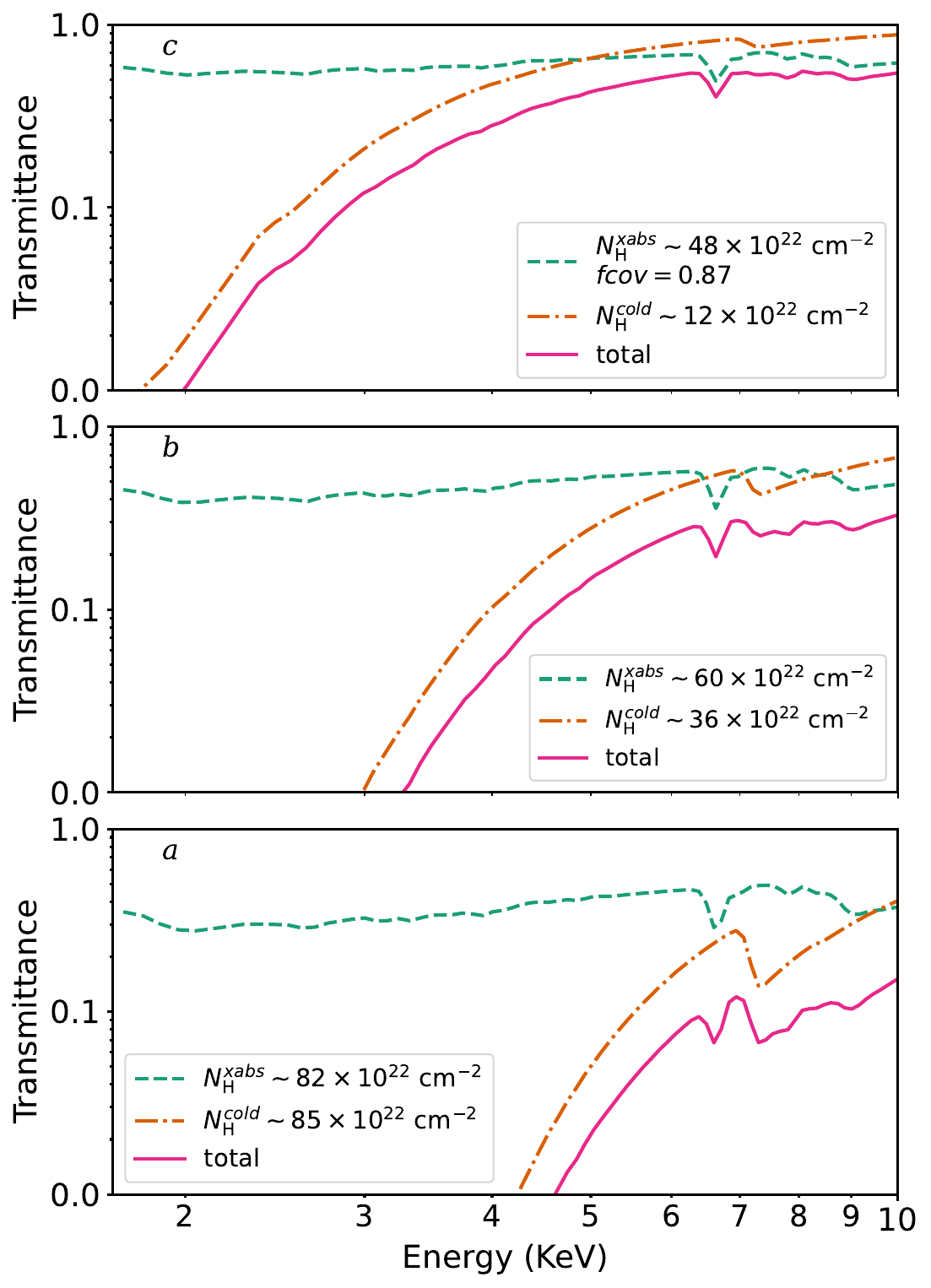}
   \caption{Transmission plots (absorbed spectrum normalized by continuum emission) for the direct emission of the flux-resolved spectra of the \textit{deep-dipping} interval of obs 401 (a,b,c). Each plot shows the transmission of the ionized absorber (\texttt{xabs}, dashed line), absorption from the total amount of cold material (dashed-dotted line), and the total transmission (solid line). All the absorption models are also convolved with the resolution of EPIC pn. The column density of the absorbers is indicated, together with the covering factor (fcov) values. If not indicated, the covering factor has a value of 1. }
   \label{fig: spec_401_tp_de}
   \end{figure}

    The spectral results of the three intensity intervals from the \textit{deep-dipping} of obs 401, including model and residuals are shown in Fig \ref{fig: spec_401_fc}. Table \ref{tab: res_401_fc} lists the corresponding best-fit parameters value. We used the same model and approach as in the time-resolved part of this work (i.e. we kept the continuum parameters coupled amongst the intervals and modeled any observed spectral changes with variations in absorbers' properties). 
    
    Cold absorption appears highest in spectrum \textit{a} with a value $N_{\mathrm{H}}$ $\sim$82$\times10^\mathrm{22}$ $\mathrm{cm}^{-2}$, almost a factor of 8 higher than in \textit{c}, and a factor 2.5 higher than \textit{b}. This intuitively suggests that in \textit{a} we sample the maximal absorption from the blocking medium.
    Concerning the ionized absorption, we could fit one absorber in each of the three spectra. The column density is $N_{\mathrm{H}}$ $\sim$82$\times10^\mathrm{22}$ $\mathrm{cm}^{-2}$ in \textit{a}, while it decreases to $\sim$60$\times10^\mathrm{22}$ $\mathrm{cm}^{-2}$ and $\sim$48$\times10^\mathrm{22}$ $\mathrm{cm}^{-2}$ in intervals \textit{b} and \textit{c} respectively.
    We coupled log$\xi$ for the three flux intervals but set an upper limit to 3.45 (the value found in \textit{clumps$\#$1}, closest in time to the \textit{deep-dipping} see Table \ref{res_401}).This was done to avoid unphysical values from the spectrum of interval \textit{a} (with the lowest statistics) to interfere with the fit, given that we found the log$\xi$ value pegging at the upper limit of any range given. However, we note that in intervals \textit{b} and \textit{c} the absorber is well constrained below the upper limit value when allowed to vary independently.
    The covering factor of the absorber is >0.57 in \textit{a} and \textit{b}, and 0.87 for \textit{c}.

    These spectral changes in column density and covering fraction values of the absorbing gas in the three intensity intervals of the \textit{deep-dipping} can be visualized with transmission plots. The transmission of each absorber is obtained by normalizing the absorbed spectrum by the continuum emission.
    In Fig \ref{fig: spec_401_tp_de}, we plot the transmission of all these gas phases (solid magenta line), the ionized gas (dashed light green line), and the cold intrinsic component (dashed-dotted orange line). The plots show how the reduction in column density of both the ionized and the cold gas between spectrum \textit{b} and \textit{c} results in more radiation passing through. This is also visible in the light curve, where interval \textit{c} shows a significant amount of the direct continuum being filtered through (Fig~\ref{fig: lc_fold_int}). 

     \begin{figure*}
   \centering
   \includegraphics[width=0.90\textwidth,angle=0]{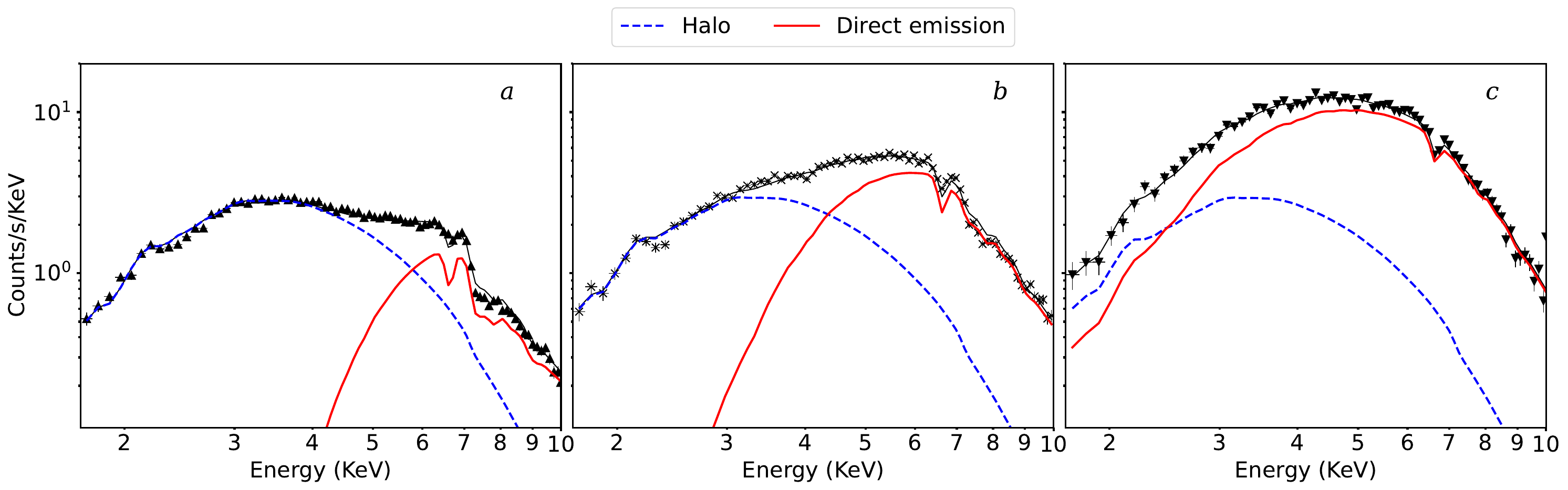}

   \caption{Display of model components for the flux-resolved spectra of the \textit{deep-dipping} interval of obs 401 (\textit{a}, \textit{b}, \textit{c}). We show how the direct and scattered emission (see Fig \ref{fig: model}) dominate for different flux levels. The direct part of the emission is shown in red (solid line) and the halo component is shown in blue (dotted line). The halo contribution in spectra \textit{b} and \textit{a} results in the double bumped/flattened shape observed.} 
   \label{fig: spec_401_fc2}
   \end{figure*} 

    \subsection{The scattering halo component}
    \label{subsub: halo comp}

    Below we describe the complexities of the treatment of the halo component. The halo model we used is described in $\S$\ref{sub: model}. While fitting, the halo column density was not kept constant throughout all intervals, as was done in \cite{DiazTrigo2006}, because our spectral analysis spans intervals larger than the halo response time. However, given that the scattering halo contribution is the most significant when the overall flux is low (i.e. \textit{deep-dipping}), during intervals of still relatively high flux (i.e. \textit{clumps$\#$1} through \textit{persistent}) the halo contribution is more difficult to constrain. Therefore, some values were forced to stay constant nonetheless.

    Specifically, for each observation, the halo column density value was obtained 
    from the \textit{clumps$\#$1} (i.e the closest one to the \textit{deep-dipping}), and then kept fixed for the other intervals (i.e. \textit{clumps$\#$2}, \textit{shallow-dipping}, and \textit{persistent}), to avoid unphysical values. This was done because the time-resolved analysis ($\S$ \ref{sub:time resolved}) did not include the \textit{deep-dipping} intervals.
    For all observations, we found $N_{\mathrm{H}}$ $\sim$10$\times10^\mathrm{22}$ $\mathrm{cm}^{-2}$, also consistent with the value of cold absorption in the \textit{persistent}, direct part, of the emission (Table \ref{fig: spec_601}, \ref{res_301}, \ref{res_401}, \ref{res_501}). 
    For the optical depth of the scattering halo, we allowed $\tau_{\mathrm{1}}$ (defined in $\S$ \ref{sub: model}) to vary between 1.8 and 2.5 according to the values found from \cite{DiazTrigo2006} and \cite{BC2000}.
    
    The halo component appears essential to the modeling of the flux-resolved spectra. 
    As shown in Fig \ref{fig: spec_401_fc2},  in the three intensity intervals sampled, the direct source emission (solid red line) and halo component (dotted blue line), contribute differently to the total flux. 
    During the maximal coverage of the \textit{deep-dipping} (spectrum \textit{a}), the direct part of the emission is extremely attenuated by the high column density of the material blocking the radiation. Most photons below 5 keV come from the scattered halo emission, resulting in the apparent bump at lower energies and the flattened shape.
    When the absorber column density decreases and some radiation can pass through the medium (spectrum \textit{b}), this recovery of direct source emission overpowers the halo contribution, resulting in the apparent flattening of the spectrum below 5 keV.
    In the moment of maximum flux recovery sampled during the \textit{deep-dipping} (spectrum \textit{c}), the direct emission dominates the observed spectral shape, resembling those shown in the time-resolved analysis.

    \section{Discussion}
    \label{sec: discussion}
    In this work, we set out to investigate the properties of the dipping plasmas in the NS LMXB \f 4U by means of a time and flux-resolved analysis of four EPIC pn \foo XMM observations.

    \subsection{The nature of the dipping material}
    \label{sub: nature of dipping discus}
   
    We combine both the flux-resolved (\textit{deep-dipping} intervals \textit{a}, \textit{b}, \textit{c}) and time-resolved (\textit{clumps\#1}, and \textit{clumps\#2}) results of obs 401 and attempt a description of the nature and structure of the material causing the dipping. The intervals mentioned have the same continuum parameters (see Table \ref{tab: res_401_fc} and \ref{res_401}), hence we can compare changes in the absorbers properties (i.e. column density and ionization parameter) relative to each interval.

    The results of our flux-resolved analysis explain the \textit{deep-dipping} modulations observed in the light curves and their corresponding spectral changes with varying column density of both cold and ionized absorbers, as well as changes in covering factor of the ionized ones. The ionization parameter of the gas remains consistent throughout the flux-selected intervals of the \textit{deep-dipping} (Table \ref{tab: res_401_fc}). 
    These results suggest that 
    the obscuring medium during the \textit{deep-dipping} interval is characterized by a multiphase (i.e. composed of both colder and ionized gas), nonuniform, and clumpy medium. This could imply that the intervals of maximal coverage (\textit{a}) are caused by an accumulation of smaller clumps, and that the fact that radiation is partially blocked (\textit{b}), or can be fully transmitted (\textit{c}), results from the nonuniform nature of the medium.

    Our time-resolved spectral analysis appears consistent with such line of reasoning. As the bulk of the bulge moves away from the line of sight, in the \textit{clumps\#1} and \textit{clumps\#2} intervals, we sample the 'leftover' substructure of the dipping plasmas. More specifically, in \textit{clumps\#1}, the column density of both ionized and cold absorbers is reduced by a factor $\sim8$ compared to the deepest dipping activity (flux interval \textit{a}). However, the ionization of the gas remains log$\xi\sim$3.4, which hints at a single type of dipping plasma changing in column density. This suggests that the modulation observed in \textit{clumps\#1} could be the result of a smaller aggregation of clumps compared to the \textit{deep-dipping} interval \textit{a}. In \textit{clumps\#2}, the column densities are reduced even further, which is expected given that the light curve only shows sporadic clumps intervening along the line of sight. In this time interval, log$\xi\sim$3.7, suggesting that the gas is more highly ionized than during \textit{clumps\#1} and the \textit{deep-dipping} flux-selected intervals. This is reasonable, as a smaller column density of the medium (i.e. smaller amount of clumps) would allow for more radiation to pass through and ionize the gas. Finally, when the dipping plasma moves completely out of the line of sight, eventually, no more clumps are observed in the light curves (e.g.  \textit{shallow-dipping} and \textit{persistent} intervals).

    As we have shown that results are consistent among the four observations, we find that the picture of a multiphased and clumpy medium of similar nature that is changing in column density holds consistently for all our dipping data set.

    Finally, we can make some simple estimates of the bulge's and its clumps' physical sizes in relation to the system. The radius of the accretion disk is estimated to be $\sim$1.1$\times10^{6}$ $\mathrm{km}$ \citep[by][calculated from the Roche Lobe size of the neutron star]{BC2004}.
    The \textit{deep-dipping} interval lasts $\sim$~2.1 hours (about 10\% of the orbital period, Fig \ref{fig: lc_fold_int}). This results in a physical size of $\sim$6$\times10^{5}$ $\mathrm{km}$ along the circumference, assuming that the bulge region is located at the rim of the disk. Picturing the bulge as an agglomeration of smaller clumps, we could interpret the modulations observed in \textit{clumps\#2} as a ``unit scale'' for the clumps, and perform a similar calculation. From the light curves, we see that the modulation that resembles a ``single'' clump (Fig \ref{fig: lc_fold_int}, second panel, orange interval) occurs in $\sim$5$\times10^{-4}$ fraction of the orbital period, which corresponds to $\sim$38~$\mathrm{s}$ and a physical size of $\sim$940 $\mathrm{km}$. This estimated value is comparable with the predicted one by eq. 9 in \cite{Frank1987} - obtained by assuming the neutron star mass to be 1 M$_{\odot}$, the accretion stream half-width to be $\sim$10$^{9}$ cm, and the residual function \textit{f} to be 6 (i.e. that 60$\%$ of the accretion rate remains in the residual stream).

    We note that clumps of shorter duration are also present, however they appear much lower in column density and must therefore have a smaller impact on the obscuration due to the bulge. The ratio of the two physical sizes (the bulge extension along the disk rim and size of the ``single'' clump) gives $\sim$7$\times10^{3}$ as an estimate of the number of clumps that could make up the bulge. Our estimate is significantly lower than the predicted number of clumps at a given time according to eq. 10 in \cite{Frank1987}. This is to be expected, since we only estimate the number of clumps present in the bulge's core region (i.e. \textit{deep-dipping}), and do not account for any of its outer layers or residual components (i.e. \textit{clumps\#1} and \textit{clumps\#2}). Furthermore, our estimate likely represents a lower limit, given that we are assuming all of the obscuring structure to be at the edge of the disk, while some of it could likely extend to the inner regions of the disk \citep{Frank1987}.

    \subsection{An extended accretion disk atmosphere}
    \label{subs: ext atmosph}

    Following our additional analysis (Appendix \ref{app: B}), both \textit{persistent} and \textit{shallow-dipping} intervals show evidence for the possible presence of ionized gas with column density $N_{\mathrm{H}}$ $\sim$3--4 $\times10^\mathrm{22}$ $\mathrm{cm}^{-2}$ and $N_{\mathrm{H}}$ $\sim$1-2 $\times10^\mathrm{22}$ $\mathrm{cm}^{-2}$, and ionization log$\xi\sim$ 4.7 and log$\xi\sim$ 3.9, respectively (see Table \ref{tab: pers_shall_table}).

     With regard to the \textit{persistent} intervals, \cite{Xiang2009} had detected an absorbing component with log$\xi\sim$4.3 in a \textit{Chandra} observation, all throughout intervals close and far from dipping \citep['near dip' and 'far dip' from Fig 1 in][]{Xiang2009}.  We find an ionized absorber consistent with the \textit{Chandra} study during \textit{persistent} intervals, but are unable to detect the ionized atmosphere during intervals showing dipping or clumps, likely due to the limited spectral resolution of our data.  
    In $\S$ \ref{sub: nature of dipping discus}, we estimated the physical size of the ``unit'' clumps we observe (i.e. modulations in \textit{clumps\#2}), as well as the accretion disk radius. Assuming our black body temperature is mostly representative of the disk, we can get an estimate of the disk scale height to be H $\sim$ 6$\times10^{6}$ $\mathrm{km}$ (assuming a neutron star mass of $\sim$1 M$_{\odot}$). If we assume the unit clumps to be spherical, the size of the clumps is significantly smaller than the disk scale height. This implies the possibility of detecting two ionized absorbers, the atmosphere and the clumps. 
    
    With regard to the \textit{shallow-dipping} intervals, the fact that we are able to detect an ionized absorber regardless of whether the shallow modulation is evident or not from the light curves (Table \ref{tab: pers_shall_table}), could hint either at a type of modulation that is always present, but that we are not able to resolve, or at some phase-dependent structure, perhaps unrelated to any dipping activity. 

    In addition, we believe it insightful to compare our findings with the \textit{Chandra} results, specifically with the second absorber found in their 'far dip' intervals \citep[Fig.~1 in][]{Xiang2009}. We notice that some of these intervals align in orbital phase with our \textit{clumps\#1}, \textit{clumps\#2}, and \textit{shallow-dipping}. The lower sensitivity of the \textit{Chandra} light curves implies the possibility that such modulations could not be resolved and, consequentially, included in the persistent analysis. 
    Our results suggest that the second component found by \textit{Chandra} (and also associated with the disk rim) could be the left over substructure of the bulge, and that it should likely not differ from the other absorber found in their 'near dip' intervals.

    An ionized atmosphere that is changing in properties could also be suitable to explain the ionized gas in the \textit{shallow-dipping} interval. Simultaneous different plasma states are to be expected \citep{DiazTrigo2016}, therefore different ionization zones and density layers may be present throughout the extended accretion disk atmosphere. 
    The presence and nature of this absorber in the \textit{shallow-dipping} remain unclear at this stage, but can be addressed in future work with higher-resolution spectral data.

    \section{Summary and conclusions}
    \label{sec: summary}
    
    In this work, we use a large \foo XMM data set, focused on the dips of \f 4U over different epochs, to characterize the structure and nature of the gas at the impact point of the gas stream onto the accretion disk.
    We studied the details of the dipping activity systematically by classifying the variety of modulations observed across the light curves (see Fig \ref{fig: lc_fold_int}). 
    
    \begin{enumerate}
    \item By combining a time-resolved and flux-resolved spectral analysis, we were able to point at different structures that contribute to the X-ray obscuration observed, as well as get insight into the nature of the dipping plasmas. Specifically:
    \begin{itemize}
        \item Moving from the most to the least dense parts of the bulge, the absorbers' column density and covering factor progressively decrease, while the ionization remains constant, hinting at the presence of clumps. With the analysis of the heaviest part of the dipping interval, we observe variations in the amount of radiation passing through, which is proof of an inhomogeneous nature. Hence, the bulge appears to be clumpy and non-uniform, as well as multiphase, as composed of both cold and highly ionized material;
        \item We made some simple estimates of the bulge' and its clumps' physical sizes. Taking the shortest clumps modulation we observe (\textit{clumps\#2}, Fig \ref{fig: lc_fold_int}) as a unit scale, the \textit{deep-dipping} modulation could be composed of at least $\sim$7$\times10^{3}$ of such clumps;
    \end{itemize}
    
    \item The additional analysis carried out for the intervals where dipping modulation was not evident or absent, led to insightful analysis related to the accretion disk structure:
    \begin{itemize}
    \item We consistently found a highly ionized atmosphere in all observations. The ionization of the gas compares with the current literature value and previous studies. Although we are not able to resolve it, in our analysis we estimate that with higher spectral resolution, this component should also be detected during moments where dipping is less prominent (e.g. \textit{clumps\#1} and \textit{clumps\#2}). Currently, it remains unclear if it would also be detected during the deepest dipping activity (i.e. \textit{deep-dipping});
    \item The nature of the \textit{shallow-dipping} modulation remains unconstrained;
    \end{itemize}

    \item With a systematic comparison of the spectral results throughout all observations, were also able to gain insight into the evolution of these dipping plasmas over time. Specifically:

    \begin{itemize}
        \item The properties of the dipping plasmas of \f 4U appear not to change in the long time scale sampled by our data ($\sim$ 6 months). It appears they could be affected by possible turbulent mixing in the impact region on shorter time scales (in between consecutive orbits). However, variability studies found that turbulence is not necessarily needed to explain the observed line width \citep{Xiang2009}, therefore, this could be further revisited with XRISM \citep{xrism_tashiro2020} data which provides higher resolution; 
    \end{itemize}

        \item Finally, we note that the dust-scattering halo is essential in modeling the source spectrum during the most heavily obscured states (the lowest count spectra, \textit{b} and \textit{a} in Fig \ref{fig: spec_401_fc}). This was also pointed out by \cite{DiazTrigo2006}; A full characterization in time of the halo was not possible with this \foo XMM dataset.
    \end{enumerate}

    To conclude, this work highlights that dipping light curves and their spectra are powerful tracers for changes in column density and properties of the gas layers in the accretion disk environment of LMXBs. With the aim of further understanding what sets the bulge properties, future work could involve looking into possible connections between the properties of the ionized gas in the disk and the system binary parameter (e.g. accretion rate, orbital period, and disk size). In addition, questions still left open can be addressed with higher resolution XRISM data.

    \vfill
    A Zenodo reproduction package for this paper is available at DOI:10.5281/zenodo.17394600 upon publication.

    \begin{acknowledgements}
    This work was based on observations obtained with of \foo XMM data, a European Space Agency (ESA) science mission with instruments and contributions directly funded by ESA Member States and the USA (NASA). The Space Research Organisation of the Netherlands is financially supported by NWO. The authors would like to thank O. Porth for insightful discussions.
    \end{acknowledgements}

\bibliographystyle{aa} 
\bibliography{ref} 

\begin{appendix}

\clearpage
\onecolumn
\section{Appendix A: Spectral fitting results for obs 301, 401, and 501}
\label{app: A}
Below we show the details of the spectral analysis of obs 301, 401, and 501, specifically their spectra and best fit spectral parameters. 

\vfill

    \begin{figure}[H]
    \centering
    \includegraphics[width=11.5cm,angle=0]{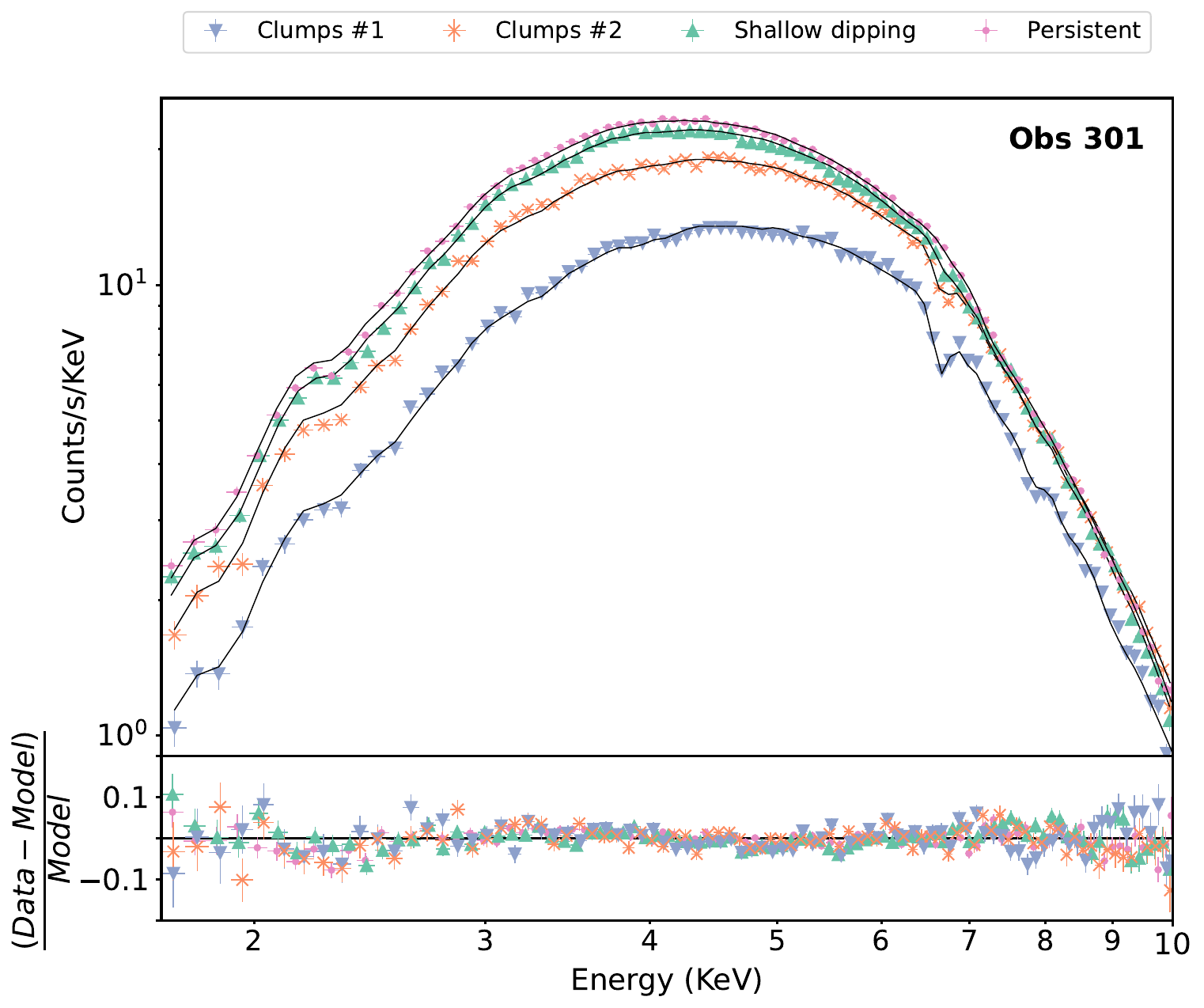}
   \caption{Spectra from the selected time intervals of obs 301 with the same color-coding as Fig \ref{fig: lc_fold_int}. Black solid lines represent the model for each spectrum, and residuals are shown in the bottom panel.}
    \label{fig: spec_301}
    \end{figure}

    \begin{table}[H]
      \caption[]{Best fit parameters and errors for the time-resolved spectral results obs 301.} 
         \label{res_301} 
        \centering
        \scalebox{0.85}{\begin{tabular}{c c c c c c c} 
            \hline
            \hline
            \noalign{\smallskip}
            \textbf{Obs 301} & & & Clumps$\#$1 & Clumps$\#$2 & Shallow & Persistent\\
            \noalign{\smallskip}
            \hline
            \noalign{\smallskip}
            \small{Comp} & \small{Param} & \small{Unit} & & & & \\
            & & & & & & \\
            \texttt{hot$_{gal}$} & $N_{\mathrm{H}}$ & \small{($10^{\mathrm{22}}$ $\mathrm{cm}^{-2}$)} & & & & 2.2 (f)\\[1ex]
            \texttt{hot$_{loc}$} & $N_{\mathrm{H}}$ & \small{($10^{\mathrm{22}}$ $\mathrm{cm}^{-2}$)} & 11.5$\pm$0.2 & 8.84$\pm0.1$ & 7.60$\pm0.03$ & 7.22$\pm0.14$ \\[1ex]
            \texttt{bb}&  norm  & \small($10^\mathrm{11}$ $\mathrm{cm}^{2}$) & & 57$_{-7}^{+9}$ & & 94$_{-12}^{+9}$ \\[1ex]
            & \textit{T} & \small{(keV)} & & 1.51$\pm0.02$ & & 1.41$\pm0.03$ \\[1ex]
            \texttt{pow} & norm & \small{($10^{44}$ ph/s/keV)} & & 71$\pm4$ & & 47.8$_{-7.3}^{+2.2}$ \\[1ex]
            & $\Gamma$ & & & 1.92$\pm0.05$ & & 1.93$_{-0.15}^{+0.02}$ \\[1ex]
            \texttt{gaus} & norm & \small{($10^{44}$ ph/s)} & & (8.0$\pm2.5$)$\times10^{-2}$ & & (5.6$_{-1.1}^{+5.8}$)$\times10^{-2}$ \\[1ex]
            & \textit{E} & \small{(keV)} & & 6.56$\pm0.04$ & & 6.54$\pm0.01$ \\[1ex]
            & FWHM & \small{(keV)} & & 0.23 (f) & & 0.23 (f)\\[1ex]
            \texttt{xabs} & $N_{\mathrm{H}}$ & \small{($10^{\mathrm{22}}$ $\mathrm{cm}^{-2}$)} & 34$\pm$0.6 & 7.8$\pm0.6$ & 4.0$_{-0.9}^{+0.4}$ & \\[1ex]
            & log $\xi$ & & 3.53$\pm$0.02 & 3.55$\pm0.02$ & 3.79$_{-0.11}^{+0.22}$ & / \\[1ex]
            & fcov & (\%) &$>$82 & $>$94 & $>$8 & \\[1ex]
            \small{\textbf{Halo}} & & & & & &\\ [0.4ex]
            \texttt{hot$_{loc}$}& $N_{\mathrm{H}}$ & \small{($10^{\mathrm{22}}$ $\mathrm{cm}^{-2}$)} & 8.2$\pm0.1$ & 8.2 (f) & 8.2 (f) & 8.2 (f) \\[1ex]
            \texttt{etau} & tau0 & & & 2.50$_{-0.02}$ & & 2.5$_{-0.1}$ \\[1ex]
            \noalign{\smallskip}
            \hline
            \noalign{\smallskip}
            $C$ (d.o.f) & 673 (346) & & & & \\
            \noalign{\smallskip}
            \hline
        \end{tabular}}
      
    \begin{minipage}{\textwidth}

    \vspace{0.1cm}


    \small Note: (f) parameter frozen to shown value; For this observation, the continuum parameters are coupled for \textit{clumps$\#$1} and \textit{clumps$\#$2}, and for \textit{shallow-dipping} and \textit{persistent}. The "/" sign indicates that the component is not included.
    \newline In the text, we refer to results on cold absorption as \texttt{hot$_{loc}$} + \texttt{hot$_{gal}$} as discussed in $\S$ \ref{sub: model}.

    \end{minipage}
   \end{table}

    \begin{figure*}
    \centering
    \includegraphics[width=11.5cm,angle=0]{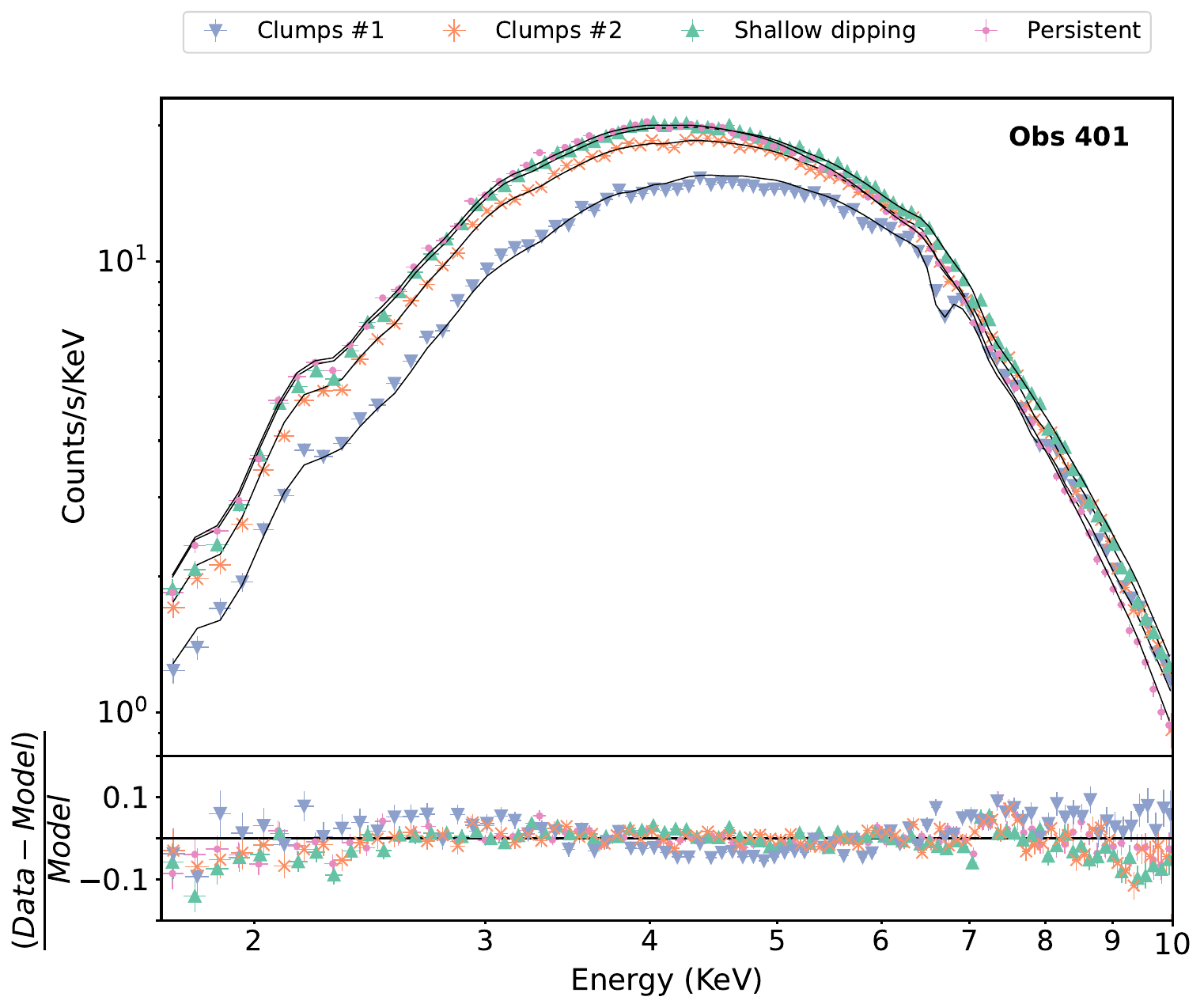}
    \caption{Spectra from the selected time intervals of obs 401 with the same color-coding as Fig \ref{fig: lc_fold_int}. Black solid lines represent the model for each spectrum, and residuals are shown in the bottom panel.}
    \label{fig: spec_401}
    \end{figure*}

    \begin{table*}
      \caption[]{Best fit parameters and errors for the time-resolved spectral results obs 401.}
         \label{res_401}
         \centering
         \scalebox{0.85}{\begin{tabular}{c c c c c c c} 
            \hline
            \hline
            \noalign{\smallskip}
            \textbf{Obs 401} & & & Clumps$\#$1 & Clumps$\#$2 & Shallow & Persistent\\
            \noalign{\smallskip}
            \hline  
            \noalign{\smallskip}
            \small{Comp} & \small{Param} & \small{Unit} & & & & \\
            & & & & & & \\
            \texttt{hot$_{gal}$} & $N_{\mathrm{H}}$ & \small{($10^{\mathrm{22}}$ $\mathrm{cm}^{-2}$)} & & & & 2.2 (f)\\[1ex]
            \texttt{hot$_{loc}$} & $N_{\mathrm{H}}$ & \small{($10^{\mathrm{22}}$ $\mathrm{cm}^{-2}$)} & 12.0$\pm0.1$ & 9.30$\pm0.04$ & 8.60$\pm0.04$ & 7.5$_{-2.3}^{+0.1}$ \\[1ex]
            \texttt{bb}&  norm  & \small($10^\mathrm{11}$ $\mathrm{cm}^{2}$) & & & 61$\pm9$ & 94$_{-12}^{+25}$\\[1ex]
            & \textit{T} & \small{(keV)} & & & 1.41$\pm0.01$ & 1.34$\pm0.04$ \\[1ex]
            \texttt{pow} & norm & \small{($10^{44}$ ph/s/keV)} & & & 73.0$\pm2.1$ & <48.0\\[1ex]
            & $\Gamma$ & & & & 1.87$\pm0.02$ & 1.97$_{-0.55}^{+0.03}$\\[1ex]
            \texttt{gaus} & norm & \small{($10^{44}$ ph/s)} & & & (7.3$\pm1.1$)$\times10^{-2}$ & (5.5$_{-4.0}^{+6.7}$)$\times10^{-2}$\\[1ex]
            & \textit{E} & \small{(keV)} & & & 6.47$\pm0.03$ & 6.42$_{-0.3}^{+1.4}$\\[1ex]
            & FWHM & \small{(keV)} & & & 0.23 (f) & 0.23 (f)\\[1ex]
            \texttt{xabs} & $N_{\mathrm{H}}$ & \small{($10^{\mathrm{22}}$ $\mathrm{cm}^{-2}$)} & 11.0$\pm0.4$ & 4.9$\pm0.4$ & & \\[1ex]
            & log $\xi$ & & 3.44$_{-0.33}^{+0.02}$ & 3.76$\pm0.06$ & / & / \\[1ex]
            & fcov & (\%) & >19 & >78 & & \\[1ex]
            & & & & & & \\
            \small{\textbf{Halo}} & & & & & &\\ [1ex]
            \texttt{hot$_{loc}$}& $N_{\mathrm{H}}$ & \small{($10^{\mathrm{22}}$ $\mathrm{cm}^{-2}$)} & 8.0 (f)& 8.0 (f) & 8.0 (f) & 8.0 (f)\\[1ex]
            \texttt{etau} & tau0 & & & & 2.50$_{-0.01}$ & 2.5
           0$_{-0.14}$\\
            \noalign{\smallskip}
            \hline
            \noalign{\smallskip}
            $C$ (d.o.f) & 810 (332) & & & & \\
            \noalign{\smallskip}
            \hline
        \end{tabular}}
    \begin{minipage}{\textwidth}

    \vspace{0.1cm}


    \small Note: (f) parameter frozen to shown value; For this observation, the continuum parameters are coupled for \textit{clumps$\#$1}, \textit{clumps$\#$2}, and \textit{shallow-dipping}. The "/" sign indicates that the component is not included.
    \newline In the text, we refer to results on cold absorption as \texttt{hot$_{loc}$} + \texttt{hot$_{gal}$} as discussed in $\S$ \ref{sub: model}.

    \end{minipage}
   \end{table*}

    \begin{figure*}
    \centering
    \includegraphics[width=11.5cm,angle=0]{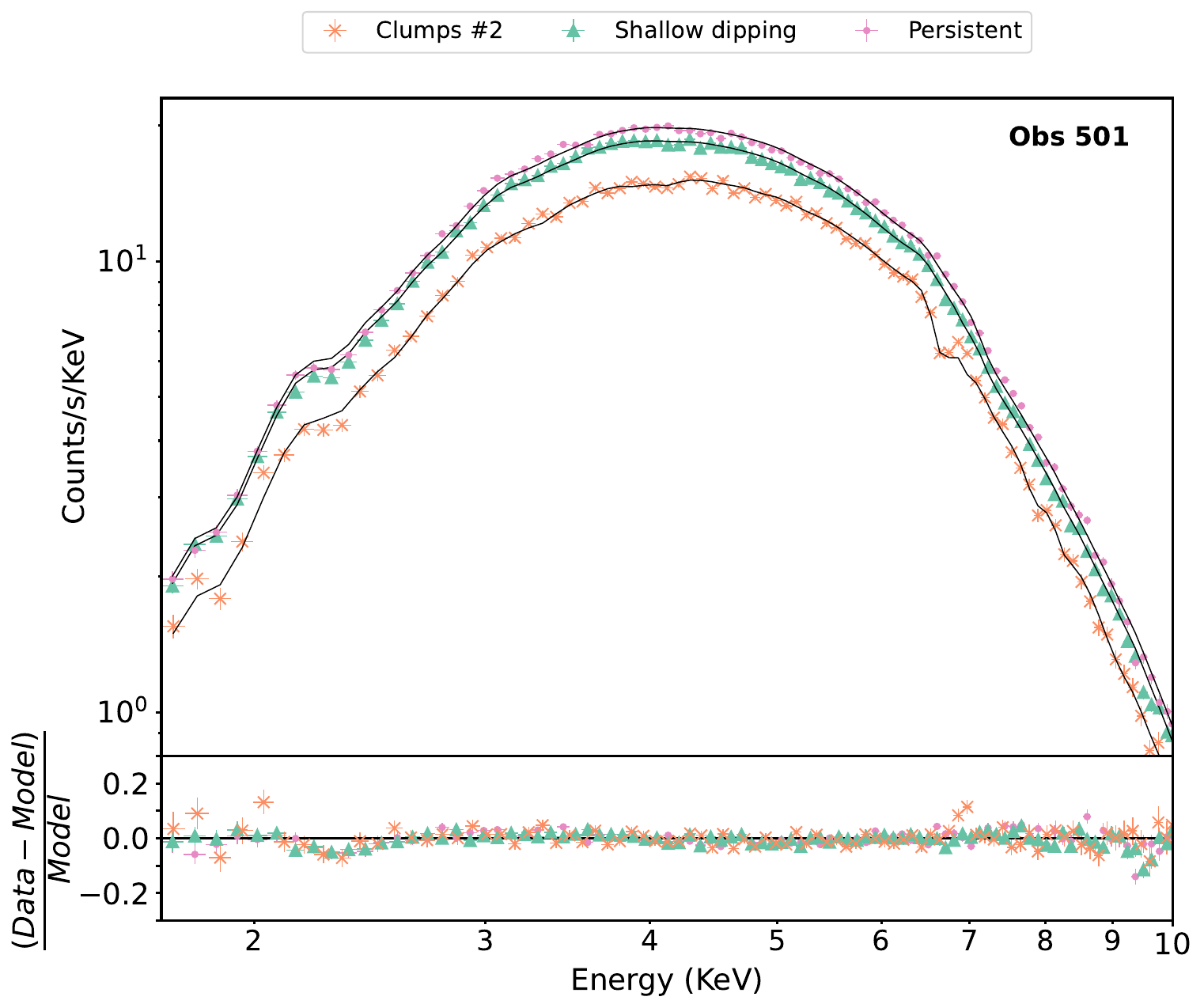}
    \caption{Spectra from the selected time intervals of obs 501 with the same color-coding as Fig \ref{fig: lc_fold_int}. Black solid lines represent the model for each spectrum, and residuals are shown in the bottom panel.}
    \label{fig: spec_501}
    \end{figure*}

    \begin{table*}
      \caption[]{Best fit parameters and errors for the time-resolved spectral results obs 501.}

         \label{res_501}
         \centering
         \scalebox{0.85}{\begin{tabular}{c c c c c c} 
            \hline
            \hline
            \noalign{\smallskip}
            \textbf{Obs 501} & &  & Clumps$\#$2 & Shallow & Persistent\\
            \noalign{\smallskip}
            \hline
            \noalign{\smallskip}
            \small{Comp} & \small{Param} & \small{Unit} & & & \\
            & & & & & \\
            \texttt{hot$_{gal}$} & $N_{\mathrm{H}}$ & \small{($10^{\mathrm{22}}$ $\mathrm{cm}^{-2}$)} & &  & 2.2 (f)\\[1ex]
            \texttt{hot$_{loc}$} & $N_{\mathrm{H}}$ & \small{($10^{\mathrm{22}}$ $\mathrm{cm}^{-2}$)} & 8.8$_{-0.1}^{+0.5}$ & 7.82$\pm0.04$ & 7.88$\pm0.03$ \\[1ex]
            \texttt{bb}&  norm  & \small($10^\mathrm{11}$ $\mathrm{cm}^{2}$) & & & 98$_{-6}^{+9}$ \\[1ex]
            & \textit{T} & \small{(keV)} & & & 1.30$\pm0.01$ \\[1ex]
            \texttt{pow} & norm & \small{($10^{44}$ ph/s/keV)} & & & 52.2$\pm1.6$ \\[1ex]
            & $\Gamma$ & & & & 1.92$\pm0.02$ \\[1ex]
            \texttt{gaus} & norm & \small{($10^{44}$ ph/s)} & & & (5.6$\pm0.9$)$\times10^{-2}$ \\[1ex]
            & \textit{E} & \small{(keV)} & & & 6.44$\pm0.04$\\[1ex]
            & FWHM & \small{(keV)} & & & 0.23 (f)\\[1ex]
            \texttt{xabs} & $N_{\mathrm{H}}$ & \small{($10^{\mathrm{22}}$ $\mathrm{cm}^{-2}$)} & 28.7$_{-0.5}^{+0.9}$ & 10.5$_{-0.2}^{+11.2}$ & \\[1ex]
            & log $\xi$ & & 3.73$\pm0.06$ & 4.85$_{-0.20}^{+0.14}$ & / \\[1ex]
            & fcov & (\%) & >38 & >9 & \\[1ex]
            & & & & & \\
            \small{\textbf{Halo}} & & & & &\\ [1ex]
            \texttt{hot$_{loc}$}& $N_{\mathrm{H}}$ & \small{($10^{\mathrm{22}}$ $\mathrm{cm}^{-2}$)} & 8.0$\pm0.1$ & 8.0 (f) & 8.0 (f) \\[1ex]
            \texttt{etau} & tau0 & & & & 2.50$_{-0.04}$ \\
            \noalign{\smallskip}
            \hline
            \noalign{\smallskip}
            $C$ (d.o.f) & 462 (260) & & & & \\
            \noalign{\smallskip}
            \hline
        \end{tabular}}
    \begin{minipage}{\textwidth}

    \vspace{0.1cm}


    \small Note: (f) parameter frozen to shown value; All parameters are coupled when no value is shown, while "/" indicates that the component is not included. 
    \newline In the text, we refer to results on cold absorption as \texttt{hot$_{loc}$} + \texttt{hot$_{gal}$} as discussed in $\S$ \ref{sub: model}.

    \end{minipage}
   \end{table*}

\clearpage
\onecolumn
\section{Additional analysis on \textit{persistent} and \textit{shallow-dipping}}
 \label{app: B}

Below we show the results of the additional analysis we carried out for \textit{persistent} and \textit{shallow-dipping}. 
For the \textit{persistent} intervals, the aim was to investigate the presence of an ionized atmosphere in the disk. For the \textit{shallow-dipping}, we wanted to investigate the properties of this modulation that seems to be the least consistent in all observations. Specifically, in obs 301 and 601 it appears as a gradual modulation of the continuum emission, while in obs 401 and 501 no evident modulation is present (Fig \ref{fig: lc_fold_int}).

    \begin{table}[H]
      \caption[]{Parameter values and errors for the additional analysis on \textit{persistent} and \textit{shallow-dipping} intervals of all four observations.}

         \label{tab: pers_shall_table}
         \centering
         \resizebox{10cm}{!}{\begin{tabular}{c c c c c c c} 
            \hline
            \hline
            \noalign{\smallskip}
            & & \textbf{301} & \textbf{401} &  \textbf{501} & \textbf{601}\\           
            \noalign{\smallskip}
            \hline
            \noalign{\smallskip}
            \textit{Persistent} & & &\\ [1ex]
            \texttt{xabs} & & &\\
            $N_{\mathrm{H}}$ & \small{($10^{\mathrm{22}}$ $\mathrm{cm}^{-2}$)} & 3.6$_{-1.6}^{+9.3}$ & 2.9$_{-2.6}^{+10.0}$ & 2.9$_{-1.5}^{+6.3}$ & 2.9$_{-2.1}^{+4.4}$\\[0.9ex]
            log $\xi$ &  & 4.7$_{-0.4}^{+0.3}$ & & \\[0.9ex]
            $C$ (d.o.f) & 707 (335) & 
            & &  & \\[1.6ex]
            \textit{Shallow-dipping} & & &\\ [1ex]
            \texttt{xabs} & & &\\
            $N_{\mathrm{H}}$ & \small{($10^{\mathrm{22}}$ $\mathrm{cm}^{-2}$)} & 1.9$_{-0.6}^{+0.9}$& 0.7$_{-0.5}^{+0.7}$& 0.9$_{-0.5}^{+0.7}$ & 1.4$_{-0.7}^{+1.1}$\\[0.9ex]
            log $\xi$ &  & 3.9$\pm0.1$ &  &  & \\[0.9ex]
            $C$ (d.o.f) & 540 (319) & 
            & &  & \\[1.4ex]
            \noalign{\smallskip}
            \hline
        \end{tabular}}
   \begin{minipage}{\textwidth}

    \vspace{0.1cm}

    \vspace{0.1cm}

    Note: The covering factor value of the absorber has value 1 for all obs. For both the \textit{shallow-dipping} and \textit{persistent} results: \newline - log $\xi$ is coupled for all observations; \newline - the continuum parameters (not shown) are within the 1$\sigma$ errors of those of the time-resolved analysis (Table \ref{res_301}, \ref{res_401}, \ref{res_501}, \ref{tab: res_601}); \newline - the halo column density value (not shown) is frozen to the previously obtained value for each observation (Table \ref{res_301}, \ref{res_401}, \ref{res_501}, \ref{tab: res_601}).
    \end{minipage}
   \end{table}

 \section{Note on continuum modeling}
    \label{subsec: model complex}
    Here we describe some of the additional complexities of our baseline model and fitting. 
    
    As mentioned in $\S$ \ref{sub: model}, in our model we assume that the accretion flow emission stays constant throughout all intervals, however, this appears not to be the case for observation 301 and 401.
    In obs 301, the \textit{clumps$\#$1} and \textit{clumps$\#$2} spectra were fitted with a continuum independent from the \textit{persistent} one (Table \ref{res_301}). The changes observed in the best-fit values are a slight decrease in the black body normalization and an increase in the power law one (Table \ref{res_301}). In obs 401, the same type of changes in continuum parameters are observed, the only difference being that the different continuum was needed for the \textit{shallow-dipping} interval as well, in addition to \textit{clumps$\#$1} and \textit{clumps$\#$2} (Table \ref{res_401}).

    We notice that these changes in continuum parameters occur at different time intervals (i.e. likely being unrelated to any source 'state' change, but rather linked to possible degeneracy in our model). Therefore we carefully tried to investigate these changes further. 

    Specifically, we first attempted to heavily rely on the continuum best fit values of the \textit{persistent} interval from each observation, as we expect this to be where the continuum would physically dominate.
    Therefore, for both obs 301 and 401, we took the best fit values of each \textit{persistent} interval, defined from these a range of $\pm$15\%, and fitted the other intervals by allowing their continuum parameters to vary within this range.
    This resulted in a worse fit with increased residuals (especially at energies >8.5 keV and <5 keV), and continuum parameters pegging at the lower/upper limit of the specified range. Finally, we also tried freeing the covering factor of the cold absorber, without any effect (i.e. covering factor remaining at a value of 1).

    We believe that a possible explanation for these observed changes in continuum could be related to our current lack of understanding of the geometry of the corona for the source. Specifically, it could be the case that one of the two continuum components could be differently absorbed than the other one. For example, the emission from the disk and the region near the neutron star (black body component) could be covered completely by the bulge, but the corona could be more vertically extended \citep[e.g. spreading layer geometry favored for the Comptonizing component in this source, see][]{Gnarini2024}. This implies that, during the different time intervals we sample, we might be seeing different amount of radiation from the corona being scattered (also given the high inclination of the system), which could result in different absorption for the two continuum components. 

    Finally, we note that the uncertainties in the modeling of the dust scattering halo ($\S$ \ref{subsub: halo comp}) for intervals where its contribution is not dominant (i.e. \textit{shallow-dipping} and \textit{persistent}) could also play a role in the changes in continuum we observe. 
    In addition, our current model does not account for scattered emission due to local cold material. Given the high column density values found ($\S$ \ref{sub:time resolved}), we expect this component to make a contribution, especially during dipping intervals. The scattering due local material would be degenerate with the halo component (i.e. they will both contribute at lower energies), hence if there is something we are incorrectly accounting for about either one, this might have an effect on the best fit values of the continuum parameters as well (e.g. normalization of the power law component). However our conclusions about the global properties of the dipping medium should not be affected.

\end{appendix}
    
\end{document}